\documentclass[10pt,aps,prc,floatfix,twocolumn,nofootinbib, superscriptaddress,preprintnumbers]{revtex4-1}
\usepackage{amsfonts,amsmath,amssymb,bm,xspace}
\usepackage{color,graphicx}
\usepackage{bbm,bm}
\usepackage{dcolumn}               
\graphicspath{{./figures/}}

\usepackage{hyperref}               

\usepackage{array,physics}
\usepackage{dcolumn}
\usepackage[normalem]{ulem}
\newcolumntype{d}[1]{D{.}{.}{#1}}

\newcommand{\fet}[1]{\mbox{\boldmath $#1$}}

\newcommand{\NLO}{\ensuremath{{\rm NLO}}\xspace}
\newcommand{\NNLO}{\ensuremath{{\rm N}{}^2{\rm LO}}\xspace}
\newcommand{\NNNLO}{\ensuremath{{\rm N}{}^3{\rm LO}}\xspace}

\hypersetup{%
    pdfsubject=Paper,
    unicode=true,
    breaklinks=true,
     colorlinks   = true,
     linkcolor = blue,
     citecolor = blue,
     menucolor = blue,
     urlcolor = blue
}

\begin{document}
\preprint{LA-UR-23-25233}
\preprint{NP3M-P2300016}

\title{Maximally local two-nucleon interactions at N$^3$LO\\ in $\Delta$-less chiral effective field theory}

\author{R. Somasundaram}
\email{rsomasun@syr.edu}
\affiliation{Theoretical Division, Los Alamos National Laboratory, Los Alamos, New Mexico 87545, USA}
\affiliation{Department of Physics, Syracuse University, Syracuse, New York 13244, USA}

\author{J.E. Lynn}
\affiliation{Theoretical Division, Los Alamos National Laboratory, Los Alamos, New Mexico 87545, USA}
\affiliation{Technische Universit\"at Darmstadt, Department of Physics, 64289 Darmstadt, Germany}
\affiliation{ExtreMe Matter Institute EMMI, GSI Helmholtzzentrum f\"ur Schwerionenforschung GmbH, 64291 Darmstadt, Germany}

\author{L. Huth}
\affiliation{Technische Universit\"at Darmstadt, Department of Physics, 64289 Darmstadt, Germany}
\affiliation{ExtreMe Matter Institute EMMI, GSI Helmholtzzentrum f\"ur Schwerionenforschung GmbH, 64291 Darmstadt, Germany}

\author{A. Schwenk}
\affiliation{Technische Universit\"at Darmstadt, Department of Physics, 64289 Darmstadt, Germany}
\affiliation{ExtreMe Matter Institute EMMI, GSI Helmholtzzentrum f\"ur Schwerionenforschung GmbH, 64291 Darmstadt, Germany}
\affiliation{Max-Planck-Institut f\"ur Kernphysik, Saupfercheckweg 1, 69117 Heidelberg, Germany}

\author{I. Tews}
\email{itews@lanl.gov}
\affiliation{Theoretical Division, Los Alamos National Laboratory, Los Alamos, New Mexico 87545, USA}

\begin{abstract}
We present new maximally local two-nucleon interactions derived in $\Delta$-less chiral effective field theory up to next-to-next-to-next-to-leading order (N$^3$LO), which include all contact and pion-exchange contributions to the nuclear Hamiltonian up to this order.
Our interactions are fit to nucleon-nucleon phase shifts using a Bayesian statistical approach, and explore a wide cutoff range from $0.6-0.9$~fm ($\approx 660 - 440$~MeV).
These interactions can be straightforwardly employed in  quantum Monte Carlo methods, such as the auxiliary field diffusion Monte Carlo method.
Together with local three-nucleon forces, calculations with these new interactions will provide improved benchmarks for the structure of atomic nuclei and serve as crucial input to analyses of astrophysical phenomena of neutron stars, such as binary neutron-star mergers.
\end{abstract}

\maketitle

\section{Introduction}
\label{sec:introduction}

Astrophysical explosions involving neutron stars (NSs), like supernovae and NS mergers, are fascinating phenomena to study nuclear physics at the extremes. 
NSs and their mergers explore nuclear matter reaching the highest densities in the cosmos, making them ideal laboratories to elucidate strong nuclear interactions.
These interactions manifest themselves in the form of the nuclear matter equation of state (EOS), which connects NSs with experiments of neutron-rich nuclei.
Recently, multi-messenger analyses, combining state-of-the-art nuclear theory with astrophysical observations, have provided a wealth of new information on the EOS~\cite{Annala:2017llu,Lim:2018bkq,Capano:2019eae,Annala:2019puf,Dietrich:2020efo,Raaijmakers:2021uju,Essick:2021kjb,Huth:2021bsp}.
These analyses used constraints on the EOS of dense matter that were obtained from many-body calculations using interactions from chiral effective field theory (EFT)~\cite{Hebeler:2009iv,Tews:2012fj,Coraggio:2014nva,Lynn:2015jua,Drischler:2017wtt}.
Chiral EFT provides a systematic expansion of nuclear interactions and is connected to the fundamental theory of strong interactions, quantum chromodynamics~\cite{Epelbaum:2008ga,Machleidt:2011zz,Hammer:2019poc}. 
While there has been a lot of progress in recent years to improve chiral EFT constraints, it is crucial to reduce uncertainties of theoretical models for nuclear interactions to fully exploit the multitude of anticipated data from NS observations and nuclear experiments in the coming years.

One way to achieve this goal is to perform calculations at higher orders in the chiral EFT expansion.
In this paper, we introduce a new family of local chiral EFT two-nucleon ($NN$) interactions at next-to-next-to-next-to-leading order (\NNNLO) that can be employed in quantum Monte Carlo (QMC) computational methods. 
QMC methods are among the most precise many-body methods~\cite{Carlson:2014vla} but they require local interactions as input, see Refs.\cite{Gezerlis:2013ipa,Gezerlis:2014zia} for details.
In the past, local interactions from chiral EFT have been developed up to next-to-next-to-leading order (\NNLO) in the $\Delta$-less approach~\cite{Gezerlis:2013ipa,Gezerlis:2014zia} and including short-range pieces at \NNNLO in the $\Delta$-full approach~\cite{Piarulli:2014bda,Piarulli:2017dwd}.
Fully local chiral interactions have also recently been developed in the $\Delta$-less approach where all available local operators up to \NNNLO were considered, even those that would be connected by antisymmetrization and the Fierz rearrangement freedom (FRF)~\cite{Saha:2022oep}.
In particular, $4$ leading-order (LO) operators, 8 local next-to-leading order (NLO) operators, and all 11 local \NNNLO operators were considered while all nonlocal operators and their associated physics were not included.

In this paper, we develop maximally local interactions for use in QMC methods while using FRF~\cite{Huth:2017wzw}. 
By maximally local, we mean a complete set of operators in which the number of nonlocal operators is as low as possible, consistent with FRF. 
Local chiral EFT interactions up to \NNLO, developed following our approach, have been used in QMC methods and provide a good description of nuclei with $A\lessapprox 20$~\cite{Lynn:2015jua,Lonardoni:2017hgs,Martin:2023dhl} and dense matter~\cite{Lynn:2015jua,Tews:2018kmu,Lonardoni:2019ypg}.
However, uncertainties in these calculations are still sizable and result from both the truncation of the chiral expansion and regulator artifacts.
These uncertainties affect QMC predictions of nuclei and the nuclear equation of state, including the nuclear symmetry energy and related observables~\cite{Lonardoni:2019ypg,Novario:2021low,Essick:2021kjb,Essick:2021ezp}.
Forthcoming astrophysics constraints from gravitational-wave (GW) observatories will provide precision data on dense matter~\cite{Finstad:2022oni} and it is key that uncertainties in many-body calculations at low densities are reduced to enable best-possible analyses of these exciting new data.
Here, we focus on developing novel maximally local \NNNLO $NN$ interactions that can be  used in QMC calculations of various nuclear systems. 
In a forthcoming paper, we will include the parameter-free \NNNLO three-nucleon ($3N$) interactions~\cite{Bernard:2007sp,Bernard:2011zr}, the charge-symmetry and charge-independence breaking corrections, and QMC calculations of many-body nuclear systems.
Calculations at higher orders in the EFT expansion are expected to reduce theoretical uncertainties by a factor of $2 - 3$~\cite{Drischler:2017wtt,Drischler:2020hwi}. 

The interactions developed here include a set of 21 contact operators, out of which 4 are nonlocal. 
All pion-exchange interactions are local and fully included.
The local interactions developed explore a wide range of cutoffs, $R_0=0.6-0.9$~fm ($\approx 660 - 440$~MeV), in order to reduce the impact of regulator artifacts. 
This is an important aspect of our calculation since previous studies have shown that regulator artifacts from local regulators are larger than for nonlocal regulators~\cite{Lynn:2015jua,Dyhdalo:2016ygz,Huth:2017wzw}.
High-cutoff interactions have smaller regulator artifacts and can easily be employed in QMC calculations, in contrast to most other many-body methods that require softer interactions for convergence.
Our N$^3$LO interactions, which we name N$^3$LO$_{\rm LA}$-09 to N$^3$LO$_{\rm LA}$-06, are fit to $NN$ scattering phase shifts using Bayesian methods.
This allows us to explicitly model EFT truncation uncertainties when performing the fits and we show how these uncertainties evolve with the cutoff $R_0$. 
We also perform least-squares fits to the NN scattering phase shifts that do not incorporate EFT truncation uncertainties. 
The comparison between the two ways of fitting indicates the importance of modeling the EFT truncation uncertainties when chiral interactions are calibrated to data. 
We demonstrate that local high-cutoff interactions perform better than their softer counterparts, in the sense that they better reproduce $NN$ scattering phase shifts and lead to smaller EFT truncation uncertainties. 
Finally, we show that although our interactions are not fit to the properties of the deuteron, our model predictions for these are in good agreement with experimental data, especially for our high-cutoff interactions.  

This paper is organized as follows. 
In Sec.~\ref{sec:Hamiltonian}, we give the explicit form of the interactions that we use in this work. 
The couplings are fit to $NN$ scattering phase shifts and the details of this fit are given in Sec.~\ref{sec:phase_shifts}, where we discuss both the Bayesian fit as well as the least-squares fit.
We also show how the $np$ phase shifts and their associated theoretical uncertainties change with increasing chiral order and varying the cutoff. 
In Sec.~\ref{sec:deuteron}, we use our interactions to study the properties of the deuteron. 
Our main conclusions and summary are presented in Sec.~\ref{sec:conclusion}.

\section{\NNNLO interactions from chiral EFT}
\label{sec:Hamiltonian}

In this section, we give the detailed expressions for the interactions along with the local regulators employed in this work. 
Since there are four nonlocal contact operators, we present the momentum-space expression for all the contacts that we use.
The pion-exchange terms and the local regulators, on the other hand, are treated in coordinate space. 

\subsection{Chiral EFT and QMC methods}

In atomic nuclei and nuclear matter below about twice the nuclear saturation density, chiral EFT is currently the main framework to describe nuclear interactions in a systematic order-by-order expansion~\cite{Epelbaum:2008ga,Machleidt:2011zz,Hammer:2019poc}.
The chiral EFT framework provides consistent $NN$, $3N$, and higher-body interactions, based on a low-momentum expansion of nuclear forces where the expansion parameter $Q$ is the ratio of a typical momentum of the system under study with respect to the breakdown scale $\Lambda_b$, which determines where chiral EFT becomes inapplicable.
The effects of high momenta, that would be resolved above the breakdown scale, are absorbed into a set of low-energy couplings, the strengths of which are adjusted to reproduce experimental data.
The advantages of chiral EFT over other approaches are that it (i) allows to quantify theoretical uncertainties~\cite{Epelbaum:2014efa,Drischler:2020yad,Drischler:2020hwi} and (ii) provides consistent $NN$ and many-body interactions, i.e., the same processes between different particles are described by the same low-energy constants (LECs) and operators.
Order by order, predictions become more accurate and precise by a factor of $2-3$ at the cost of more involved calculations.
Chiral EFT is valid for relative nucleon momenta below $\approx 500-600$~MeV, translating into densities below about twice the nuclear saturation density.

Solving the nuclear many-body problem is a challenging task that requires advanced computational tools. 
QMC methods are among the most precise nuclear many-body methods~\cite{Carlson:2014vla} and use stochastic techniques to extract ground-state properties of nuclear systems, providing quasi exact solutions with statistical uncertainties~\cite{Lonardoni:2017hgs}.
This is the main benefit over other computational methods whose additional approximations can lead to systematic uncertainties. 
However, QMC methods require local interactions\footnote{Note that small nonlocalities can be treated perturbatively.} 
as input, i.e., interactions with no derivatives acting on the wave functions, to nonperturbatively solve the nuclear many-body problem.
While chiral EFT is traditionally formulated in a nonlocal way and has been used to construct $NN$ interactions to \NNNLO and beyond~\cite{Entem:2017gor,Reinert:2017usi}, local chiral EFT interactions have been introduced only in the past decade~\cite{Gezerlis:2013ipa,Piarulli:2014bda,Saha:2022oep}. 
Local interactions have so far been developed up to \NNLO in chiral EFT~\cite{Gezerlis:2013ipa,Gezerlis:2014zia} on the same footing as nonlocal interactions.
In addition, maximally local interactions with selected \NNNLO contributions have been developed~\cite{Piarulli:2014bda} but these typically do not include \NNNLO pion-exchange contributions.
Finally, recent work saw the development of local \NNNLO interactions where all nonlocalities were neglected and replaced by local operators where possible, even when the latter are connected by FRF~\cite{Saha:2022oep}.
Here, we develop complete maximally local $NN$ interactions at \NNNLO in chiral EFT that are suited for QMC calculations, i.e., interactions that account for all short- and long-range contributions and all necessary local and nonlocal pieces up to that order, using FRF.
While FRF is violated when local regulators are applied~\cite{Huth:2017wzw,Saha:2022oep}, this effect induces regulator artifacts that take the form of interaction pieces of higher order in the EFT expansion, and hence, decrease in size.
We have shown that the most sizable regulator artifacts at LO can be absorbed by the regular interactions pieces at NLO~\cite{Huth:2017wzw}.
Similarly, artifacts at NLO and subleading artifacts at LO can be absorbed at \NNNLO.
Hence, the remaining artifacts are of order $Q^6$ and are therefore expected to be small.
Furthermore, we study interactions at large cutoffs where these regulator artifacts further decrease in size.
For these reasons, regulator artifacts in the NN sector are small in this work.

\subsection{Maximally local intactions at \NNNLO}

Chiral EFT interactions are given in terms of a momentum expansion and can be decomposed into short-range contact pieces and long-range pieces mediated by one and multiple pion exchanges,
\begin{equation}
    V^{(\nu)} = V^{(\nu)}_{\text{cont}} + V^{(\nu)}_{\pi}\,,
\end{equation}
where $\nu$ is the chiral order, indicating the power $Q^{\nu}$.

\subsubsection{Contact interactions}
\label{subsec:contacts}

Up to \NNLO, the contact interactions can be fully expressed using only local operators and the nonlocal spin-orbit interaction that can be treated by QMC methods~\cite{Gezerlis:2013ipa,Gezerlis:2014zia,Tews:2015ufa,Lynn:2015jua}.
The leading-order (LO) momentum-independent contact interactions are given by
\begin{equation}
V^{(0)}_{\text{cont}} = C_S + C_T  \fet \sigma_1\cdot \fet\sigma_2\,,
\label{eq:LO}
\end{equation}
where two out of four possible operators are chosen~\cite{Gezerlis:2014zia,Huth:2017wzw}. 
The remaining two operators are linearly dependent due to the required antisymmetry of the wave function in nuclear systems. 
In other words, one can construct an antisymmetrized potential $V_A$,
\begin{equation}
    V_A = \frac{1}{2}(V-\mathcal{A}[V]),
\end{equation}
where $\mathcal{A}$ is the exchange operator given as 
\begin{align}
    \mathcal{A}[V(\fet q, \fet k)] &= \frac{1}{4} (1 + \fet \sigma_1 \cdot \fet \sigma_2) (1 + \fet \tau_1 \cdot \fet \tau_2) \nonumber \\ &\quad V\left(\fet q \to  -2\fet k, \fet k \to - \frac{q}{2}\right).
\end{align}
By antisymmetrizing the potential in this manner, it can be shown that any two out of four operators, describing both $NN$ $S$-wave interaction channels, can be selected using FRF, see Refs.~\cite{Huth:2017wzw,Gezerlis:2013ipa,Gezerlis:2014zia} for more details. 

At NLO, the contact interaction is momentum dependent. For initial and final relative momenta $\mathbf{p}$ and $\mathbf{p}^\prime$, momentum transfer $\mathbf{q} = \mathbf{p}^\prime - \mathbf{p}$, and momentum transfer in the exchange channel $\mathbf{k}=(\mathbf{p}^\prime + \mathbf{p})/2$, the NLO interaction is given by
\begin{align}
V^{(2)}_{\text{cont}}&= C_1\, \fet q^2  + C_2\, \fet q^2 \fet \tau_1\cdot \fet\tau_2 + C_3\, \fet q^2 \fet \sigma_1\cdot \fet \sigma_2 \label{eq:NLO} \nonumber \\ &\quad + C_4\, \fet q^2 \fet \sigma_1\cdot \fet\sigma_2 \fet \tau_1\cdot \fet\tau_2  +\frac{i}{2} C_5\  (\fet \sigma_1+ \fet\sigma_2)\cdot(\fet q \times \fet k) \nonumber \\ &\quad + C_6  \, (\fet \sigma_1\cdot \fet q) \,(\fet\sigma_2 \cdot \fet q)   + C_7   \, (\fet \sigma_1\cdot \fet q) \,(\fet\sigma_2 \cdot \fet q )\fet \tau_1\cdot \fet\tau_2,
\end{align}
where, again, a subset of 7 independent out of 14 possible operators has been chosen. 
At \NLO, the operators are selected such that the interaction is fully local (i.e., independent of $\mathbf{k}$) except for the spin-orbit interaction. 
Other choices are possible, too, which lead to nonlocal or partially local interactions~\cite{Entem:2003ft,Epelbaum:2004fk,Ekstrom:2015rta,Entem:2017gor,Reinert:2017usi}.
There are no new contact operators that appear at \NNLO.

We now turn to \NNNLO where there are a total of 30 possible contact operators~\cite{Epelbaum:2004fk},
\begin{widetext}
\begin{align}
V^{(4)}_{\text{cont}}&= \alpha_1\, \fet q^4  + \alpha_2\, \fet q^4 \fet \tau_1\cdot \fet\tau_2 + \alpha_3\, \fet q^4 \fet \sigma_1\cdot \fet \sigma_2 + \alpha_4\, \fet q^4 \fet \sigma_1\cdot \fet\sigma_2 \fet \tau_1\cdot \fet\tau_2 \nonumber \\
&\quad  +\alpha_5\, \fet k^4 + \alpha_6\, \fet k^4 \fet \tau_1\cdot \fet\tau_2  +\alpha_7\, \fet k^4 \fet \sigma_1\cdot \fet\sigma_2 + \alpha_8\, \fet k^4 \fet \sigma_1\cdot \fet\sigma_2 \fet \tau_1\cdot \fet\tau_2 \nonumber \\
&\quad +\alpha_9\, \fet q^2 \fet k^2 + \alpha_{10}\, \fet q^2 \fet k^2 \fet \tau_1\cdot \fet\tau_2 +\alpha_{11}\, \fet q^2 \fet k^2 \fet \sigma_1\cdot \fet\sigma_2 + \alpha_{12}\, \fet q^2 \fet k^2 \fet \sigma_1\cdot \fet\sigma_2 \fet \tau_1\cdot \fet\tau_2 \nonumber\\
&\quad +\alpha_{13}\, (\fet q \times \fet k)^2 \nonumber + \alpha_{14}\, (\fet q \times \fet k)^2 \fet \tau_1\cdot \fet\tau_2 +\alpha_{15}\, (\fet q \times \fet k)^2 \fet \sigma_1\cdot \fet\sigma_2 + \alpha_{16}\, (\fet q \times \fet k)^2 \fet \sigma_1\cdot \fet\sigma_2 \fet \tau_1\cdot \fet\tau_2 \nonumber \\
&\quad +\frac{i}{2} \alpha_{17}\, \fet q^2  (\fet \sigma_1+ \fet\sigma_2)\cdot(\fet q \times \fet k)  +\frac{i}{2} \alpha_{18}\, \fet q^2  (\fet \sigma_1+ \fet\sigma_2)\cdot(\fet q \times \fet k)\fet \tau_1\cdot \fet\tau_2 \nonumber \\
&\quad +\frac{i}{2} \alpha_{19}\, \fet k^2  (\fet \sigma_1+ \fet\sigma_2)\cdot(\fet q \times \fet k)  +\frac{i}{2} \alpha_{20}\, \fet k^2  (\fet \sigma_1+ \fet\sigma_2)\cdot(\fet q \times \fet k)\fet \tau_1\cdot \fet\tau_2 \nonumber \\
&\quad +\alpha_{21}\, \fet q^2 \, \fet \sigma_1\cdot \fet q \,\fet\sigma_2 \cdot \fet q  \nonumber +\alpha_{22}\, \fet q^2 \, \fet \sigma_1\cdot \fet q \,\fet\sigma_2 \cdot \fet q \, \fet \tau_1\cdot \fet\tau_2 +\alpha_{23}\, \fet k^2 \, \fet \sigma_1\cdot \fet q \, \fet\sigma_2 \cdot \fet q  +\alpha_{24}\, \fet k^2 \, \fet \sigma_1\cdot \fet q \,\fet\sigma_2 \cdot \fet q \, \fet \tau_1\cdot \fet\tau_2  \nonumber \\
&\quad +\alpha_{25}\, \fet q^2 \, \fet \sigma_1\cdot \fet k\, \fet\sigma_2 \cdot \fet k   +\alpha_{26}\, \fet q^2 \, \fet \sigma_1\cdot \fet k \, \fet\sigma_2 \cdot \fet k \, \fet \tau_1\cdot \fet\tau_2  +\alpha_{27}\, \fet k^2 \, \fet \sigma_1\cdot \fet k \, \fet\sigma_2 \cdot \fet k   +\alpha_{28}\, \fet k^2 \, \fet \sigma_1\cdot \fet k\,  \fet\sigma_2 \cdot \fet k \, \fet \tau_1\cdot \fet\tau_2  \nonumber \\
&\quad +\alpha_{29}\,  \fet \sigma_1\cdot (\fet q \times \fet k) \, \fet\sigma_2 \cdot (\fet q \times \fet k) +\alpha_{30}\, \fet \sigma_1\cdot (\fet q \times \fet k) \, \fet\sigma_2 \cdot (\fet q \times \fet k)  \fet \tau_1\cdot \fet\tau_2 \,.
\label{eq:N3LOfullset}
\end{align}
\end{widetext}
Note, that terms $\sim (\fet{q}\cdot \fet{k})^2$ can be expressed in terms of $\fet{q}^2\fet{k}^2$ and $(\fet q \times \fet k)^2$, and hence, do not have to be included explicitly.
Also, the last two operators are related to a quadratic spin-orbit operator and the angular momentum operator by 
\begin{align}
    \frac12 \left((\fet \sigma_1 + \fet \sigma_2)\cdot (\fet q \times \fet k) \right)^2 &= \fet \sigma_1\cdot (\fet q \times \fet k) \, \fet\sigma_2 \cdot (\fet q \times \fet k) \nonumber \\
    & \quad + (\fet q \times \fet k)^2\,.
\end{align}

Due to FRF, we can again choose a subset of 15 independent operators from this over-complete set. 
When selecting the subset, however, it is crucial to choose operators in such a way that upon antisymmetrization the complete operator set is recovered.
Hence, for example, a nonlocal tensor operator cannot be replaced by a central local interaction piece as physics information at that order would be lost. One possible way to choose an independent set of operators would be to drop all terms in Eq.~\eqref{eq:N3LOfullset} that contain the product of isospin matrices $\fet\tau_1\cdot \fet\tau_2$, as done in Ref.~\cite{Epelbaum:2004fk}.
Alternatively, in this work, we choose the following subset of operators:
\begin{align}
V^{(4)}_{\text{cont}}&= D_1\, \fet q^4  + D_2\, \fet q^4 \fet \tau_1\cdot \fet\tau_2 + D_3\, \fet q^4 \fet \sigma_1\cdot \fet \sigma_2  \nonumber \\
&\quad  + D_4\, \fet q^4 \fet \sigma_1\cdot \fet\sigma_2 \fet \tau_1\cdot \fet\tau_2 + \frac{i}{2} D_5\, \fet q^2  (\fet \sigma_1+ \fet\sigma_2)\cdot(\fet q \times \fet k)  \nonumber \\
&\quad  +\frac{i}{2} D_6\, \fet q^2  (\fet \sigma_1+ \fet\sigma_2)\cdot(\fet q \times \fet k)\fet \tau_1\cdot \fet\tau_2 \nonumber \\
&\quad +D_7\, \fet q^2 \, \fet \sigma_1\cdot \fet q \,\fet\sigma_2 \cdot \fet q  \nonumber +D_8\, \fet q^2 \, \fet \sigma_1\cdot \fet q \,\fet\sigma_2 \cdot \fet q \, \fet \tau_1\cdot \fet\tau_2 \nonumber \\
&\quad  + D_9\, \fet q^2 \fet k^2 + D_{10}\, \fet q^2 \fet k^2 \fet \tau_1\cdot \fet\tau_2 +D_{11}\, (\fet q \times \fet k)^2 \nonumber \\
&\quad  + D_{12}\, (\fet q \times \fet k)^2 \fet \tau_1\cdot \fet\tau_2  +D_{13}\, \fet k^2 \, \fet \sigma_1\cdot \fet q \, \fet\sigma_2 \cdot \fet q  \nonumber \\
&\quad +D_{14}\, \fet k^2 \, \fet \sigma_1\cdot \fet q \,\fet\sigma_2 \cdot \fet q \, \fet \tau_1\cdot \fet\tau_2 \nonumber \\
&\quad +D_{15}\,  \fet \sigma_1\cdot (\fet q \times \fet k) \, \fet\sigma_2 \cdot (\fet q \times \fet k)\,.
\label{eq:n3lo}
\end{align}
The operators $D_1$ to $D_8$ are local and $D_9$ to $D_{15}$ are nonlocal. Note that the terms $D_5$ and $D_6$, while being nonlocal, can be treated by Monte Carlo methods since the momentum $\fet k$ appears only linearly. Hence we group them together with the local operators at $\NNNLO$, just as we have grouped $C_5$ along with the other 7 local operators at $\NNLO$.

We can further reduce the number of nonlocal contacts as it has been found that there are redundancies among the 15 contact operators at \NNNLO~\cite{Reinert:2017usi,Wesolowski:2018lzj}.
By performing a unitary transformation (UT) on the Hamiltonian, we can decrease the number of independent nonlocal contacts from 7 to 4~\cite{Reinert:2017usi}. 
Following Ref.~\cite{Reinert:2017usi}, we consider the unitary operator
\begin{equation}
    U = e^{\gamma_1 T_1 + \gamma_2 T_2 + \gamma_3 T_3  }\,,
\end{equation}
where $T_i$ are the three anti-Hermitian generators of the UT and $\gamma_i$ are the corresponding transformation angles. 
Similarly to Ref.~\cite{Reinert:2017usi}, we choose the following generators:
\begin{align}
    T_1 &=\frac{m_N}{2 \Lambda_b^4} \fet k \cdot \fet q \,, \label{T1}\\ 
    T_2 &= \frac{m_N}{2 \Lambda_b^4} \fet k \cdot \fet q \ \fet \tau_1\cdot \fet\tau_2 \,, \label{T2}\\
    T_3 &= \frac{m_N}{2 \Lambda_b^4} ( \fet \sigma_1\cdot \fet k \fet \sigma_2\cdot \fet q + \fet \sigma_1\cdot \fet q \fet \sigma_2\cdot \fet k ) \ \fet \tau_1\cdot \fet\tau_2 \label{T3}\,.
\end{align}
Note that Eqs.~\eqref{T1},~\eqref{T2}, and \eqref{T3} constitute one choice of basis for the UT and other choices are possible due to FRF. 
For example, Ref.~\cite{Reinert:2017usi} replaces the operator $\fet \tau_1\cdot \fet\tau_2 $ by $\fet \sigma_1\cdot \fet\sigma_2$ in Eq.~\eqref{T2} and, $\fet \tau_1\cdot \fet\tau_2 $ with $\openone$ in Eq.~\eqref{T3}. Moreover, apart from the freedom in choosing the generator basis, there are only three two-nucleon contact operators that are anti-Hermitian and Galilean invariant at order $Q^2$. Therefore, the unitary operator $U$ considered in this work represents all possible UTs that can be generated at order $Q^2$.  
We need to apply the UT only to the LO Hamiltonian since the UT, when applied to the higher-order interactions, will induce terms at order $Q^6$ and above which is beyond the desired accuracy for this work. 
The shift in the LO Hamiltonian is given by
\begin{align}
    \delta H^0 &=  U^\dagger  H^0 U - H^0 \nonumber \\
    &= \sum_i \gamma_i [H^0,T_i] + \dots \nonumber \\
    &=  \sum_i \gamma_i [(H_{\text{kin}}+ V^{(0)}_{1\pi}+V^{(0)}_{\text{cont}}),T_i] + \dots, \label{eq:ShiftHamiltonian}
\end{align}
where the dots represent terms beyond the order of our calculation. 
It can be shown that $[ V^{(0)}_{1\pi},T_i]$ and $[ V^{(0)}_{\text{cont}},T_i]$ only induce shifts to contact operators of order $Q^0$ and $Q^2$ and therefore do not need to be considered explicitly~\cite{Reinert:2017usi}.
On the other hand, the commutator with the kinetic energy generates order $Q^4$ terms (see Ref.~\cite{Reinert:2017usi}):
\begin{align}
     \sum_i \gamma_i &[H_{\text{kin}},T_i ] =   \frac{\gamma_1}{\Lambda_b^4} (\fet k \cdot \fet q )^2  + \frac{\gamma_2}{\Lambda_b^4} (\fet k \cdot \fet q )^2 \fet \tau_1\cdot \fet\tau_2 \nonumber \\ &\quad +  \frac{\gamma_3}{\Lambda_b^4} (\fet k \cdot \fet q )  ( \fet \sigma_1\cdot \fet k \fet \sigma_2\cdot \fet q + \fet \sigma_1\cdot \fet q \fet \sigma_2\cdot \fet k )  \fet \tau_1\cdot \fet\tau_2\,. \label{commutator}
\end{align}
Using the identity
\begin{equation}
    (\fet k \cdot \fet q )^2 = \fet q^2 \fet k^2 - (\fet q \times \fet k)^2\,,
\end{equation}
and Eq.~(10) of Ref.~\cite{Reinert:2017usi},
\begin{align}
    (\fet k \cdot \fet q )  ( \fet \sigma_1\cdot \fet k \fet \sigma_2\cdot \fet q &+ \fet \sigma_1\cdot \fet q \fet \sigma_2\cdot \fet k )  =  - (\fet q \times \fet k)^2 \fet \sigma_1\cdot \fet \sigma_2  \nonumber \\ &\quad + \fet q^2 \fet \sigma_1\cdot \fet k \fet \sigma_2\cdot \fet k   \nonumber \\ &\quad + \fet k^2 \fet \sigma_1\cdot \fet q \fet \sigma_2\cdot \fet q   \nonumber \\ &\quad + \fet \sigma_1\cdot (\fet q \times \fet k) \, \fet\sigma_2 \cdot (\fet q \times \fet k)\,,
\end{align}
we can express Eqs.~\eqref{eq:ShiftHamiltonian} and~\eqref{commutator} as
\begin{align}
        \delta H^0 &=  \frac{\gamma_1}{\Lambda_b^4} \fet q^2 \fet k^2  + \frac{\gamma_2}{\Lambda_b^4} \fet q^2 \fet k^2 \fet \tau_1\cdot \fet\tau_2 \nonumber \\ 
        &- \frac{\gamma_1}{\Lambda_b^4} (\fet q \times \fet k)^2 - \frac{\gamma_2}{\Lambda_b^4} (\fet q \times \fet k)^2 \fet \tau_1\cdot \fet\tau_2 \nonumber \\ 
        & - \frac{\gamma_3}{\Lambda_b^4} (\fet q \times \fet k)^2 \fet \sigma_1\cdot \fet \sigma_2 \fet \tau_1\cdot \fet\tau_2 \nonumber \\
        &+ \frac{\gamma_3}{\Lambda_b^4}   \fet q^2 \fet \sigma_1\cdot \fet k \fet \sigma_2\cdot \fet k \fet \tau_1\cdot \fet\tau_2 \nonumber \\ 
        & + \frac{\gamma_3}{\Lambda_b^4}   \fet k^2 \fet \sigma_1\cdot \fet q \fet \sigma_2\cdot \fet q \fet \tau_1\cdot \fet\tau_2 \nonumber \\
        &+  \frac{\gamma_3}{\Lambda_b^4}  \fet \sigma_1\cdot (\fet q \times \fet k) \, \fet\sigma_2 \cdot (\fet q \times \fet k) \fet \tau_1\cdot \fet\tau_2\,.
\end{align} 
Note that some of the new operators, 
\begin{align}
    \{(&\fet q \times \fet k)^2 \fet \sigma_1\cdot \fet \sigma_2 \fet \tau_1\cdot \fet\tau_2, \fet q^2 \fet \sigma_1\cdot \fet k \fet \sigma_2\cdot \fet k \fet \tau_1\cdot \fet\tau_2, \nonumber \\
        & \fet \sigma_1\cdot (\fet q \times \fet k) \, \fet\sigma_2 \cdot (\fet q \times \fet k) \fet \tau_1\cdot \fet\tau_2 \}\,,
    \label{extra}
\end{align}
are not explicitly chosen in Eq.~\eqref{eq:n3lo}, but they are linearly dependent operators; see Eq.~\eqref{eq:N3LOfullset}.
Therefore, we do not need to consider them explicitly any further. 

Having carried out the UT and including the shift to the Hamiltonian, the N$^3$LO contact interaction is now
\begin{align}
V^{(4)}_{\text{cont}}&= D_1\, \fet q^4  + D_2\, \fet q^4 \fet \tau_1\cdot \fet\tau_2 + D_3\, \fet q^4 \fet \sigma_1\cdot \fet \sigma_2  \nonumber \\
&\quad + D_4\, \fet q^4 \fet \sigma_1\cdot \fet\sigma_2 \fet \tau_1\cdot \fet\tau_2+ \frac{i}{2} D_5\, \fet q^2  (\fet \sigma_1+ \fet\sigma_2)\cdot(\fet q \times \fet k)  \nonumber \\
&\quad +\frac{i}{2} D_6\, \fet q^2  (\fet \sigma_1+ \fet\sigma_2)\cdot(\fet q \times \fet k)\fet \tau_1\cdot \fet\tau_2 \nonumber \\
&\quad +D_7\, \fet q^2 \, \fet \sigma_1\cdot \fet q \,\fet\sigma_2 \cdot \fet q  \nonumber +D_8\, \fet q^2 \, \fet \sigma_1\cdot \fet q \,\fet\sigma_2 \cdot \fet q \, \fet \tau_1\cdot \fet\tau_2 \nonumber \\
&\quad  + 
\bigg(D_9 + \frac{\gamma_1}{\Lambda_b^4} \bigg)\, \fet q^2 \fet k^2 + \bigg(D_{10} + \frac{\gamma_2}{\Lambda_b^4} \bigg)\, \fet q^2 \fet k^2 \fet \tau_1\cdot \fet\tau_2 \nonumber \\
&\quad +\bigg(D_{11} - \frac{\gamma_1}{\Lambda_b^4} \bigg)\, (\fet q \times \fet k)^2 \nonumber \\
&\quad + \bigg(D_{12} - \frac{\gamma_2}{\Lambda_b^4} \bigg)\, (\fet q \times \fet k)^2 \fet \tau_1\cdot \fet\tau_2 \nonumber \\
&\quad +D_{13}\, \fet k^2 \, \fet \sigma_1\cdot \fet q \, \fet\sigma_2 \cdot \fet q  \nonumber \\
&\quad +\bigg(D_{14} + \frac{\gamma_3}{\Lambda_b^4} \bigg)\, \fet k^2 \, \fet \sigma_1\cdot \fet q \,\fet\sigma_2 \cdot \fet q \, \fet \tau_1\cdot \fet\tau_2 \nonumber \\
&\quad +D_{15}\,  \fet \sigma_1\cdot (\fet q \times \fet k) \, \fet\sigma_2 \cdot (\fet q \times \fet k)  \,.
\end{align}
The variables $\gamma_1$, $\gamma_2$ and $\gamma_3$ are completely arbitrary parameters of the UT and can be chosen to remove nonlocal operators.
The parameter $\gamma_1$ can be chosen to be either $-\Lambda_b^4 D_9$ or $\Lambda_b^4 D_{11}$, removing either of the two corresponding contact operators.
Similarly, $\gamma_2$ can be chosen as either $-\Lambda_b^4 D_{10}$ or $\Lambda_b^4 D_{12}$. 
Finally, $\gamma_3$ can be set to $-\Lambda_b^4 D_{14}.$\footnote{In fact, due to the Fierz ambiguity, $\gamma_3$ can be used to remove also other nonlocal operators; see Eq.~\eqref{extra}.}  
Therefore, we see that using the UT we can remove 3, leaving only $4$ nonlocal operators. 

\begin{figure*}
    \centering
    \includegraphics[scale=0.5]{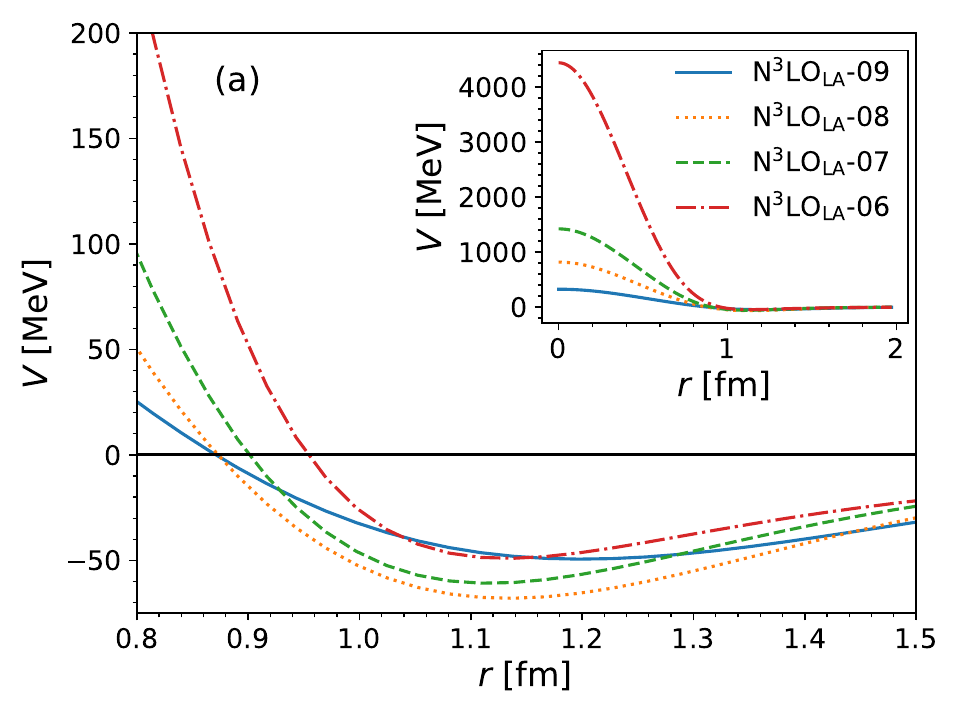}
    \includegraphics[scale=0.5]{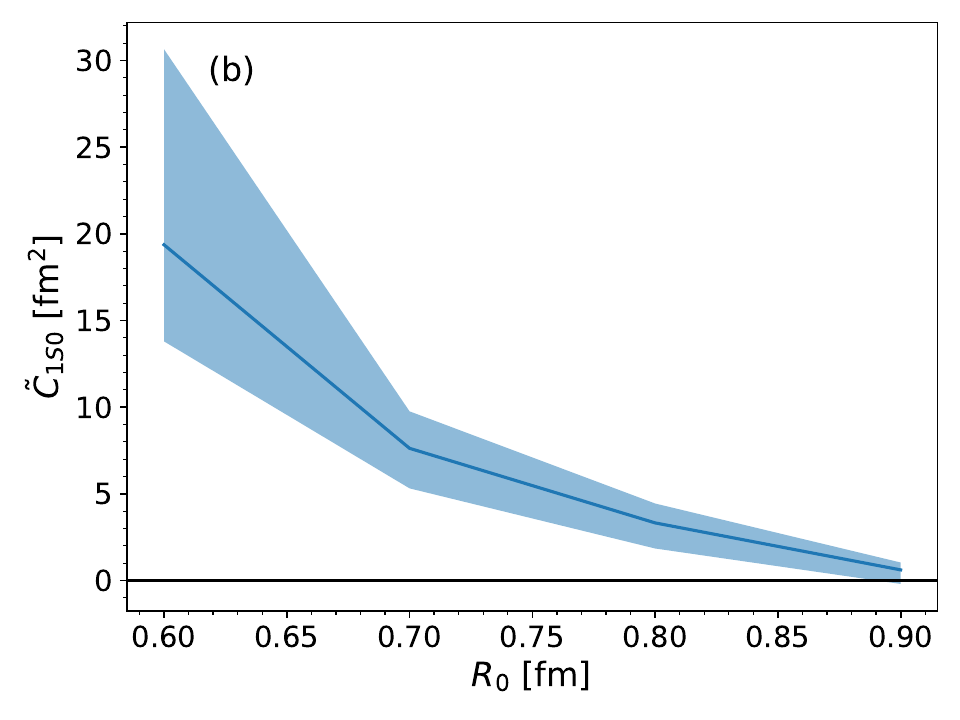}
    \caption{(a): The local part of our N$^3$LO$_{\mathrm{LA}}$ interactions in the $^1S_0$ channel (i.e., 17 local contacts and all the pion-exchange pieces) for the least-squares fit as a function of particle separation $r$. The LECs are obtained via least-squares fits as described in Sec.~\ref{sec:phase_shifts}.
    We show the interactions for different cutoff values, i.e., N$^3$LO$_{\mathrm{LA}}$-09 to N$^3$LO$_{\mathrm{LA}}$-06. (b): Cutoff dependence of the LO spectral LEC in the $^1S_0$ partial wave at \NNNLO. The solid blue line shows the least-squares results whereas the band represents the 95\% confidence level (CL) calculated using a Bayesian analysis; see Sec.~\ref{sec:phase_shifts}.}
    \label{fig:pot}
\end{figure*}

\begin{figure*}
    \centering
    \includegraphics[scale=0.34]{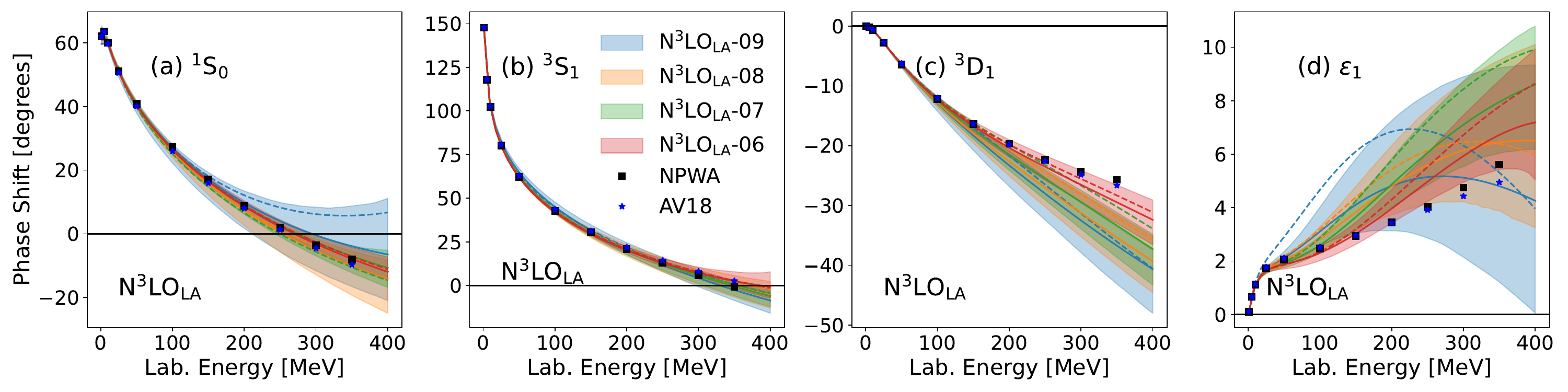}
    \includegraphics[scale=0.34]{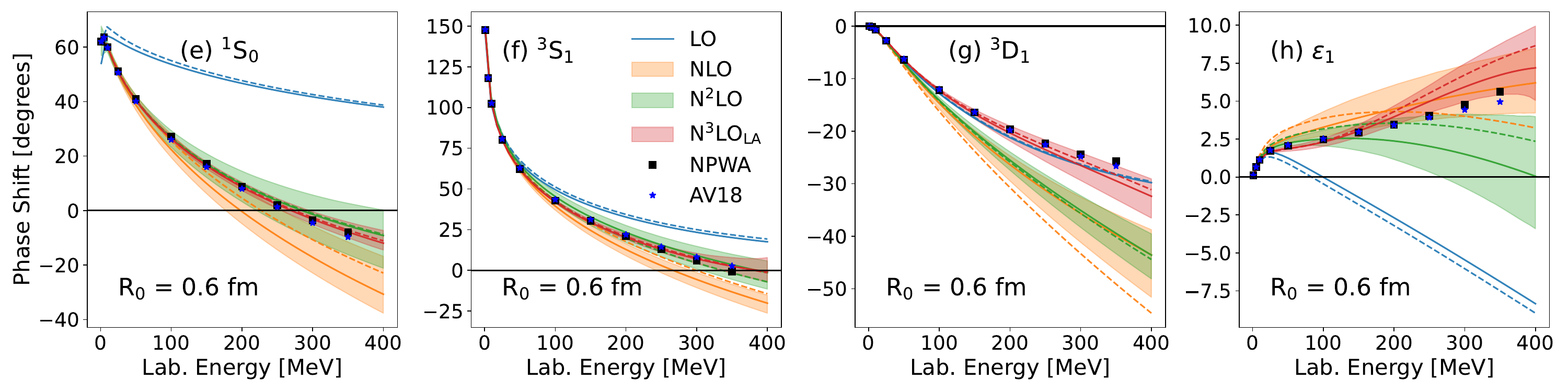}
    \includegraphics[scale=0.34]{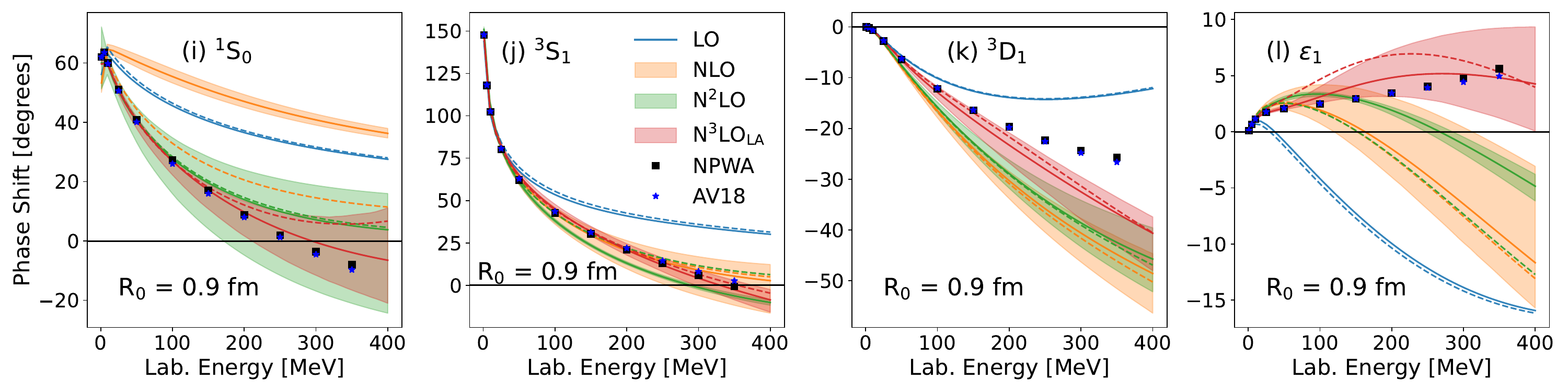}
    \caption{Phase shifts in the $^1S_0$ and $^3S_1$-$^3D_1$ partial waves.
    In panels (a)-(d), the chiral EFT order is fixed to be \NNNLO and we show results for different cutoffs as indicated in the legend.
    In panels (e)-(h) [panels (i)-(l)], we show results for different chiral EFT orders at $R_0=0.6$~fm [$R_0=0.9$~fm].
    The bands correspond to the 95\% CL. 
    The solid lines represent the solutions that maximize the Bayesian posterior distributions while the dashed lines represent the least-squares fit results.
    For comparison, we show the Nijmegen partial-wave analysis (NPWA)~\cite{Stoks:1993tb} (black squares) and the phase shifts of the AV18 potential~\cite{Wiringa:1994wb} (blue stars), which has been used in past QMC calculations.}
    \label{fig:ps_S}
\end{figure*}

In this paper, we set $\gamma_1 = - \Lambda_b^4 D_9$, $\gamma_2 = - \Lambda_b^4 D_{10}$ and $\gamma_3 = - \Lambda_b^4 D_{14}$. The nonlocal part of the N$^3$LO contact interactions then consist of only 4 operators:
\begin{align}
V^{(4)}_{\text{cont,nonlocal}}&= 
\Tilde{D}_{11}\, \fet L^2  + \Tilde{D}_{12}\, \fet L^2 \fet \tau_1\cdot \fet\tau_2 \nonumber \\
&\quad +\Tilde{D}_{13}\, \fet k^2 \, \fet \sigma_1\cdot \fet q \, \fet\sigma_2 \cdot \fet q  \nonumber \\
&\quad +\Tilde{D}_{15}\, (\fet \sigma_1 \cdot \fet L) (\fet \sigma_2 \cdot \fet L)\,,
\label{eq:nonlocal_ops}
\end{align}
where we have used the relative orbital angular momentum operator $\fet L = (\fet q \times \fet k)$.
We will refer to our maximally local \NNNLO interactions with this choice of 4 nonlocal operators by  N$^3$LO$_{\rm LA}$-09, N$^3$LO$_{\rm LA}$-08, N$^3$LO$_{\rm LA}$-07, and N$^3$LO$_{\rm LA}$-06, where the number refers to the cutoff in coordinate space. The coordinate space expressions of the \NNNLO contact operators are given in Appendix~\ref{sec:FTs} and the matrix elements of these operators in the partial wave basis is given in Appendix~\ref{sec:PWB}.
We have also considered other possible choices for the set of 4 nonlocal operators—see Appendix~\ref{sec:Different_nonlocal_sets}—but found the set chosen in this work to be
best suited for QMC methods because three operators can be directly mapped into the 18 operator channels of the phenomenological Argonne V18 (AV18) interaction~\cite{Wiringa:1994wb} that has been used extensively in QMC simulations.

In summary, our N$^3$LO$_\mathrm{LA}$ interactions contain 21 contacts in total, out of which 4 are nonlocal. 
The corresponding 21 LECs are determined by fits to $np$ phase shifts, see Sec.~\ref{sec:phase_shifts} for details.
In future work, we will also include the isospin-breaking contributions to the EFT interaction, following the strategy of Refs.~\cite{Gezerlis:2013ipa,Gezerlis:2014zia} to calculate neutron matter for the physical scattering length at all orders.

\subsubsection{Pion-exchange interactions}
\label{sec:pions}

The long-range and intermediate-range parts of chiral EFT interactions are mediated by pion exchanges. 
All pion-exchange interactions to \NNNLO are either fully local or accompanied by the spin-orbit operator, and thus, we directly give the coordinate space expressions here.

Without loss of generality, the pion-exchange contributions can be decomposed as~\cite{Epelbaum:2004fk,Saha:2022oep}
\begin{align}
    V_\pi &= V_C(r) + W_C(r)  \fet \tau_1\cdot \fet\tau_2 \nonumber \\
&\quad + (V_S(r)  +W_S(r) \fet \tau_1\cdot \fet\tau_2  ) \, \fet \sigma_1\cdot \fet\sigma_2 
\nonumber \\
&\quad +(V_T(r)  +W_T(r) \fet \tau_1\cdot \fet\tau_2  ) \,  S_{12} \nonumber \\
&\quad +(V_{LS}(r)  +W_{LS}(r) \fet \tau_1\cdot \fet\tau_2  ) \, \fet L\cdot \fet S\,,
\end{align}
where $S_{12} = (3 \fet \sigma_1\cdot \hat{\fet r} \, \fet \sigma_2\cdot \hat{\fet r} -  \fet \sigma_1\cdot \fet\sigma_2 )$ is the tensor operator and $\fet S=({\fet \sigma_1+ \fet\sigma_2})/2$ is the total spin operator.

At LO, only the one-pion exchange (OPE) contributes. It is given by~\cite{Gezerlis:2013ipa,Gezerlis:2014zia,Saha:2022oep}
\begin{align}
    W_S(r) &= \frac{g_A^2  m_\pi^2}{48 \pi F_\pi^2} \frac{e^{-x}}{r} \,, \\
    W_T(r) &= \frac{g_A^2 }{48 \pi F_\pi^2} \frac{e^{-x}}{r^3} (3+3x+x^2) \,
\end{align}
where $x = m_\pi r$, $m_\pi$ is the pion mass, $g_A$ is the axial-vector coupling constant, and $F_\pi$ is the pion decay constant. For the constants $m_\pi$, $F_\pi$, $g_A$, and the nucleon mass $M_N$, we use the same values as in Ref.~\cite{Gezerlis:2014zia}. 
Here, we use the charge-independence breaking form of the OPE as we have done before; see Ref.~\cite{Gezerlis:2014zia} for details.

At \NLO and beyond, $V_\pi$ receives contribution from two-pion exchange (TPE) diagrams. 
For these TPE pieces, we employ the expressions using spectral-function regularization (SFR)~\cite{Gezerlis:2013ipa,Saha:2022oep,Epelbaum:2004fk}.
Under this representation, the TPE potential is written in terms of spectral functions as 
\begin{align}
    V_C(r) &= \frac{1}{2\pi^2 r} \int_{2 m_\pi}^{\tilde{\Lambda}} d\mu \mu e^{-\mu r} \text{Im} V_C(i\mu) \,, \\
    V_S(r) &= -\frac{1}{6\pi^2 r} \int_{2 m_\pi}^{\tilde{\Lambda}} d\mu \mu e^{-\mu r} [\mu^2 \text{Im} \,V_T(i\mu) \\
    & \hspace{100pt} - 3\text{Im} \, V_S(i\mu)] \,, \\
    V_T(r) &= -\frac{1}{6\pi^2 r^3} \int_{2 m_\pi}^{\tilde{\Lambda}} d\mu \mu e^{-\mu r} (3+3\mu r+ \mu^2 r^2) \nonumber\\
    & \hspace{100pt} \times \text{Im} \,V_T(i\mu)
    \,, \\
    V_{LS}(r) &= \frac{1}{2\pi^2 r^3} \int_{2 m_\pi}^{\tilde{\Lambda}} d\mu \mu e^{-\mu r} (1+\mu r) \text{Im} V_{LS}(i\mu) \,,
\end{align}
and similarly for $W_{C,S,T,LS}$. 
All TPE contributions to the Hamiltonian at \NLO, \NNLO, and \NNNLO can therefore be specified in terms of the spectral functions $V_{C,S,T,LS}(i\mu)$. 
The explicit expressions for these spectral functions can be found in Appendix A of Ref.~\cite{Saha:2022oep}.
We note that we add to $V^{(4)}_\pi$ the $1/M_N$ corrections to the \NNLO TPE potential. 
These corrections nominally appear at fifth order in chiral EFT but, as is common practice~\cite{Saha:2022oep,Epelbaum:2004fk}, we include it at \NNNLO in order to arrive at an improved intermediate-range attraction. 
Also, note that unlike Ref.~\cite{Saha:2022oep}, we do not take the limit $\tilde{\Lambda} \rightarrow \infty$ but leave the SFR cutoff finite. 
In this work, the SFR cutoff $\tilde{\Lambda}$ is fixed at the value of 1~GeV, similarly to Refs.~\cite{Gezerlis:2013ipa,Epelbaum:2004fk}, and we have verified that changing it to $\tilde{\Lambda}=2$ GeV does not significantly affect our results; see Appendix~\ref{sec:SFR}. 

The strength of the TPE potential is determined by the $\pi N$ couplings which are constrained by chiral symmetry and can be determined by analyses of low-energy $\pi N$-scattering. 
Here, we employ the values obtained from a Roy-Steiner (RS) analysis of $\pi N$ scattering~\cite{Hoferichter:2015hva} at \NNNLO. 
At \NNLO, we chose the same values as in Ref.~\cite{Gezerlis:2013ipa,Gezerlis:2014zia} for consistency with our previous interactions. 
These values are very close to the values extracted from the RS analysis at \NNLO except for $c_4$.
The values of the couplings employed in this work are given in Table~\ref{tab:pi_N_couplings}.

\begin{table}
\centering
\tabcolsep=0.1cm
\def\arraystretch{1.5}
\begin{tabular}{ccc}
\hline\hline
$\pi N$ coupling                & \NNLO & \NNNLO \\
\hline
$c_1 \mathrm{[GeV}^{-1}]$        & $-0.81$ & $-1.07$  \\
$c_2 \mathrm{[GeV}^{-1}]$      & -- & 3.20 \\
$c_3 \mathrm{[GeV}^{-1}]$      & $-3.4$ & $-5.32$ \\
$c_4 \mathrm{[GeV}^{-1}]$      & 3.4 & 3.56 \\
$\Bar{d_1}+\Bar{d_2} \mathrm{[GeV}^{-2}]$      & -- & 1.04 \\
$\Bar{d_3} \mathrm{[GeV}^{-2}]$      & -- & $-0.48$ \\
$\Bar{d_5} \mathrm{[GeV}^{-2}]$      & -- & 0.14 \\
$\Bar{d}_{14}-\Bar{d}_{15} \mathrm{[GeV}^{-2}]$      & -- & $-1.90$ \\
\hline\hline
\end{tabular}
\caption{$\pi N$ couplings used in this work. 
The couplings at \NNLO are taken from Refs.~\cite{Gezerlis:2013ipa,Gezerlis:2014zia} whereas the couplings at \NNNLO are taken from Ref.~\cite{Hoferichter:2015hva}.}
\label{tab:pi_N_couplings}
\end{table}

\subsection{Regulators}
\label{sec:regulators}

Chiral EFT interactions need to be regulated at short distances and/or high momenta: 
\begin{align}
    V^{(\nu)}_{\text{cont}}  &\longrightarrow \quad V^{(\nu)}_{\text{cont}}  \times   f_{\text{short}}(r)\,, \\
     V^{(\nu)}_{\pi}  &\longrightarrow \quad V^{(\nu)}_{\pi} \times  f_{\text{long}}(r)\,.
\end{align}
To define maximally local interactions, we choose regulators that are fully local. 
We choose Gaussian regulators in position space, that can be expressed as
\begin{align}
f_{\text{short}}(r)&= \frac{n}{4 \pi R_0^3 \Gamma(3/n) }\exp\left(-\left(\frac{r}{R_0} \right)^n \right)\,,\\
f_{\text{long}}(r)&=\left( 1-\exp\left(-\left(\frac{r}{R_0}\right)^{n_1}  \right) \right)^{n_2}\,,
\end{align}
with $n=n_1=2$, $n_2=6$. 
We therefore, have one position-space cutoff $R_0$ that regulates both the short- and long-range pieces of the interaction. 
Upon Fourier transformation (FT) of the regulator functions, the position-space cutoff can be related to the momentum-space cutoff $\Lambda_c$ as $R_0 = 2/\Lambda_c$.
In this paper, we study four different cutoffs, $R_0=0.9$, $0.8$, $0.7$, and $0.6$~fm, leading to four different interactions, N$^3$LO$_{\rm LA}$-09 to N$^3$LO$_{\rm LA}$-06.

\section{Analysis of NN scattering}
\label{sec:phase_shifts}

\begin{figure*}
    \centering
    \includegraphics[scale=0.31]{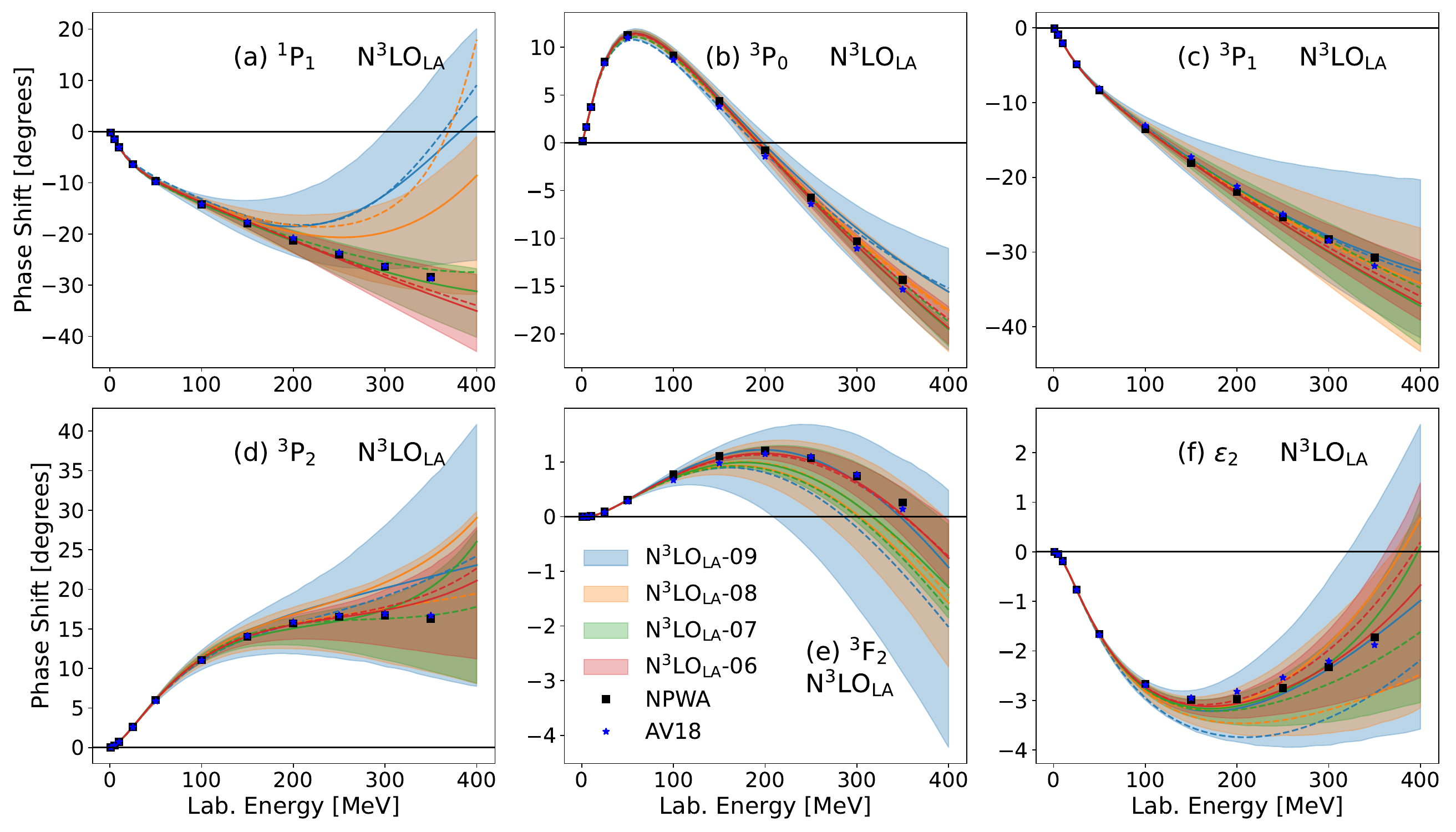}
    \includegraphics[scale=0.31]{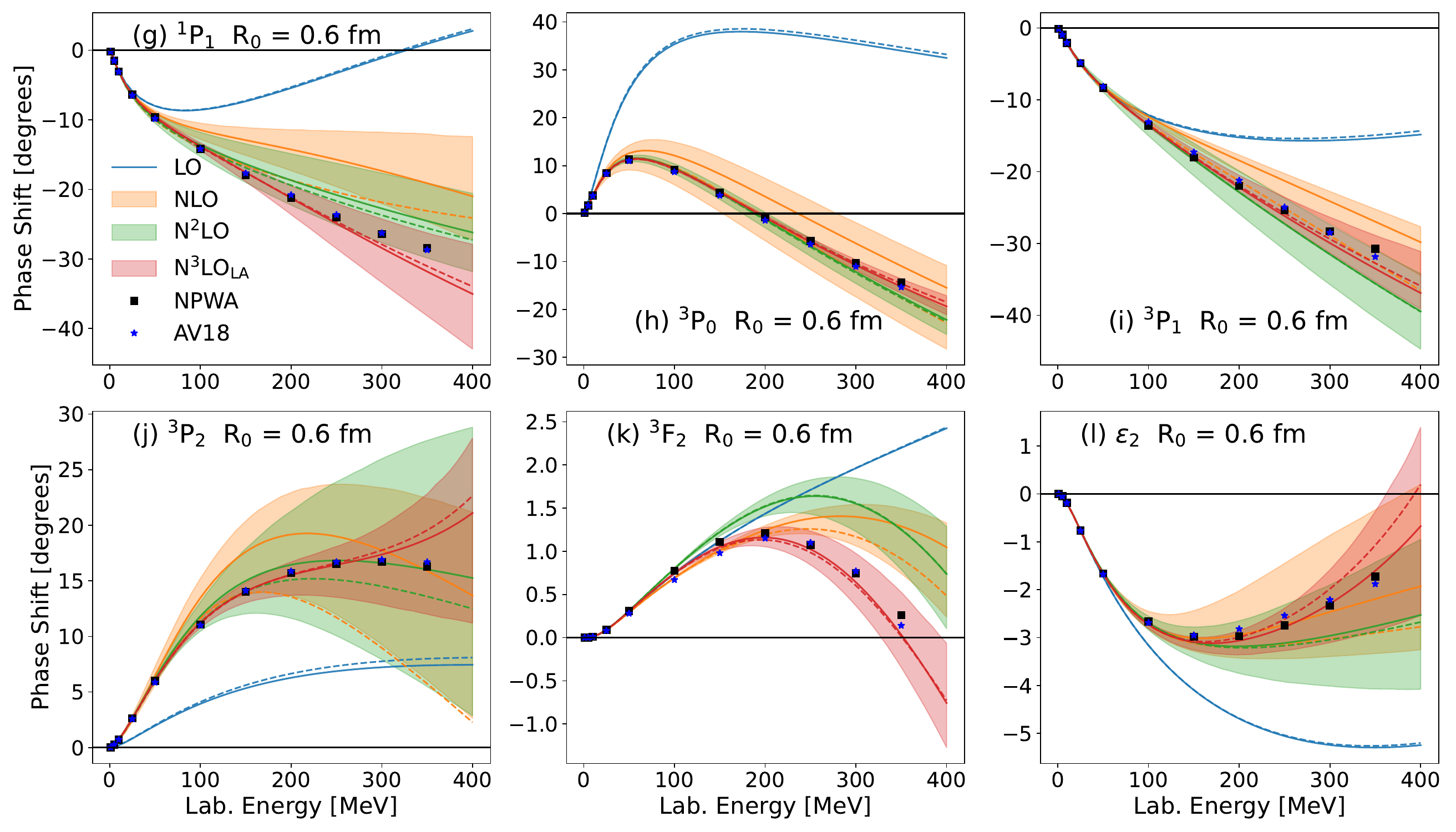}
    \includegraphics[scale=0.31]{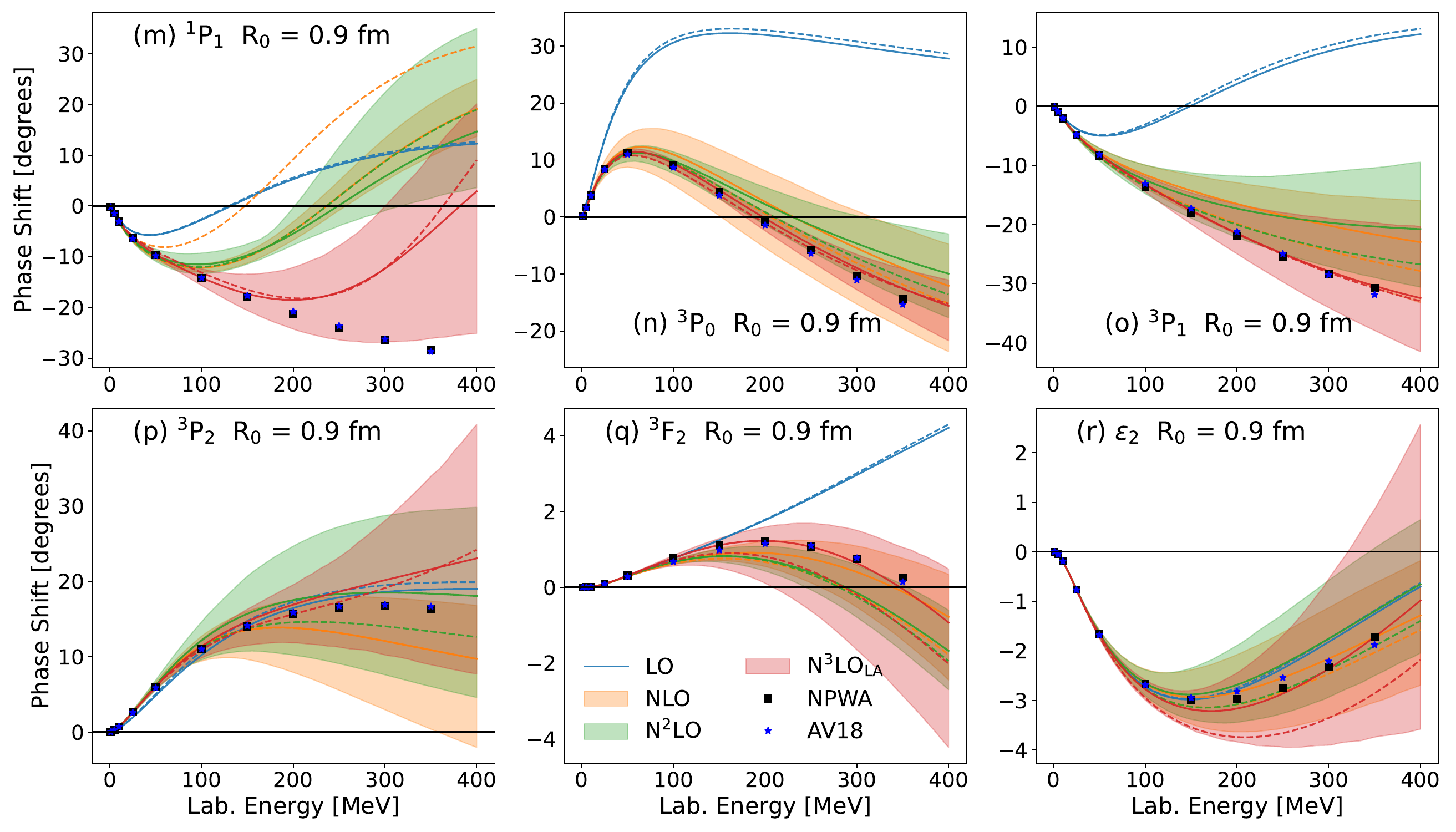}
    \caption{Same as Fig.~\ref{fig:ps_S}, but for the $^1P_1$, $^3P_0$, $^3P_1$ and $^3P_2$-$^3F_2$ partial waves.}
    \label{fig:ps_P}
\end{figure*}

In this work, the $21$ operator LECs are determined by fits to $np$ phase shifts, while in the future we will explore fits to scattering data. 
We perform the fits directly in momentum space using the formalism developed in Ref.~\cite{Huth:2018nle}.
This allows us to fit interactions that include nonlocal pieces, which is crucial at \NNNLO.
We take the phase-shift values from the Nijmegen partial-wave analysis (NPWA)~\cite{Stoks:1993tb}, and incorporate EFT truncation uncertainties by performing Bayesian fits to these data.
Bayes' theorem defines the posterior $P$ as 
\begin{equation}
    P = \frac{\mathcal{L} \times \Pi} {Z}\,,
    \label{eq:bayes}
\end{equation}
where $\Pi$ is the prior distribution on the LECs, $\mathcal{L}$ is the likelihood function that incorporates information from the phase shifts, and the normalization constant, i.e., the evidence $Z$ can, in principle, be used to perform model comparison, but we will not pursue this here.
We take the prior $\Pi$ to be uniform everywhere in parameter space, because the LECs can vary strongly in size as we vary the cutoff.
We require our likelihood function to incorporate a model for the EFT truncation uncertainties. In order to do so, we assume that the EFT expansion holds for any scattering observable $X(p)$, i.e., 
\begin{equation}
    X(p) = X_\text{ref}(p) \sum_{n=0}^{k} c_n Q^n(p)\,,
\end{equation}
where the expansion parameter $Q\equiv \frac{\text{max}(m_\pi,p)}{\text{min}(\Lambda_b,\Lambda_c)}$. 
Here, $p$ is the relative momentum, $\Lambda_b$ is the breakdown scale which we take to be $600$~MeV~\cite{Epelbaum:2014efa,Reinert:2017usi}, and $\Lambda_c$ is the momentum-space cutoff. Note that in the Weinberg power counting scheme~\cite{Epelbaum:2004fk}, $c_1 = 0$. 
For sufficiently large $k$, one might expect that the uncertainty stemming from the truncation of the EFT expansion at order $k$ can be approximated by the contribution to the observable $X$ at order $k+1$, i.e.,
\begin{equation}
    \Delta X^k(p) \approx X_\text{ref} \, Q^{k+1} c_{k+1} \,.
    \label{eq:approx}
\end{equation}
However, since $c_{k+1}$ is unknown, Epelbaum, Krebs, and Mei{\ss}ner (EKM)~\cite{Epelbaum:2014efa} proposed approximating Eq.~\eqref{eq:approx} as 
\begin{equation}
    \Delta X^{k}_{\text{EKM}} = X_\text{ref} \, Q^{j} \text{max} \big(  |c_0|, |c_1| \dots , |c_k| \big) \,,
    \label{eq:EKM}
\end{equation}
where $j = \text{max}(2,k+1)$. 
Note that this expression provides an estimate of the systematic truncation uncertainty but it does not have a strict probabilistic interpretation. 
However, it makes minimal assumptions regarding the nature of $c_n$ and the underlying probability distribution from which the $c_n$ are drawn. 
On the other hand, if one assumes that the expansion coefficients are drawn from a Gaussian distribution, the BUQEYE collaboration showed that the EFT truncation uncertainty, when summed to infinite order, also follows a Gaussian distribution whose variance is given as~\cite{Wesolowski:2018lzj}
\begin{equation}
    \sigma^2_\text{theo,BUQEYE} = \Bar{c}^2   X_\text{ref}^2 \frac{Q^{2(k+1)}}{1-Q^2} \,, 
    \label{eq:buqeye}
\end{equation}
where $\Bar{c}^2$ is the root mean square of the expansion coefficients $c_n$ up to order k. 
Equation~\eqref{eq:buqeye} holds in the limit where  all correlations across different kinematic points are ignored, see Ref.~\cite{Wesolowski:2018lzj} for the corresponding expression in the fully correlated limit. 
Assuming that experimental uncertainties are also Gaussian distributed, the BUQEYE approach leads to a Gaussian model for the likelihood function,
\begin{equation}
    \mathcal{L}_\text{BUQEYE} \propto \prod_{i} \exp \bigg\{ -\frac{1}{2} \bigg( \frac{X^{\text{exp}}_i  -X^{\text{theo}}_i}{\sigma_{i,\text{BUQEYE}}} \bigg)^2 \bigg\},
    \label{eq:buqeye_L}
\end{equation}
where $\sigma^2_{i,\text{BUQEYE}} = \sigma_\text{i,exp}^2 + \sigma_\text{i,theo,BUQEYE}^2$. 
The product over $i$ indicates a product over all considered partial waves and kinematic variables (laboratory energies).
The variable $X$ denotes an observable which, in this case, is the phase shift for a given laboratory energy and partial wave.
This likelihood function, along with priors that enforce naturalness of the LECs, was extensively used in Bayesian analyses of phase shifts in Ref.~\cite{Wesolowski:2018lzj}. 

\begin{figure*}[t]
    \centering
    \includegraphics[scale=0.37]{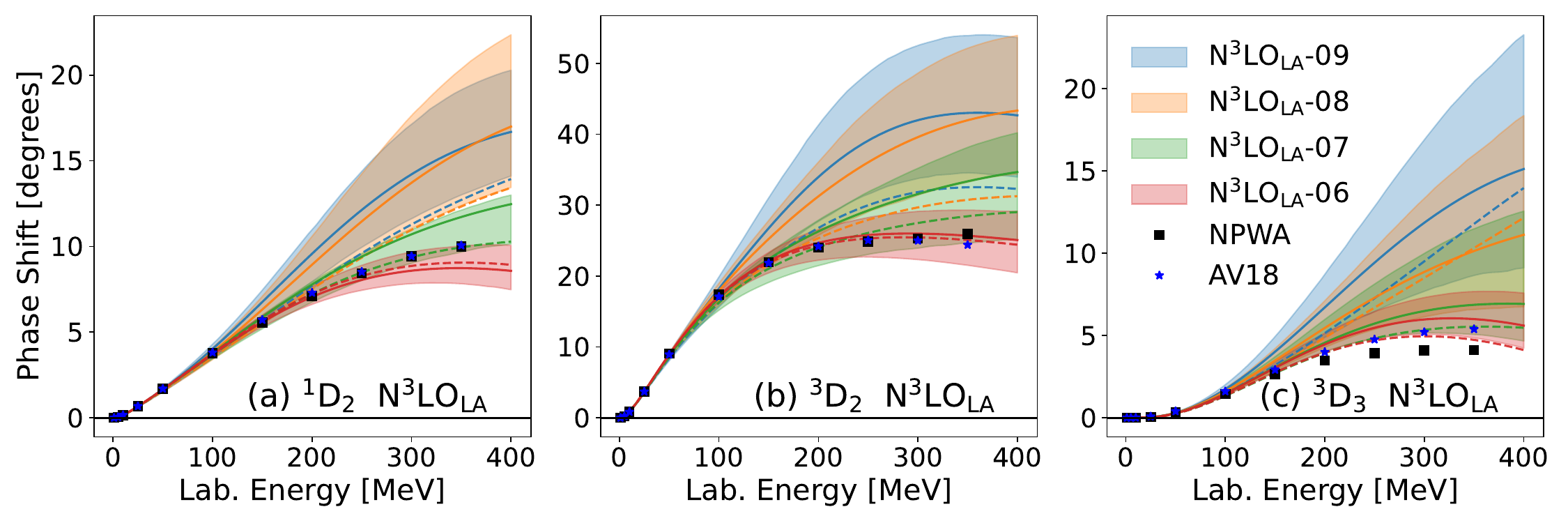}
    \includegraphics[scale=0.37]{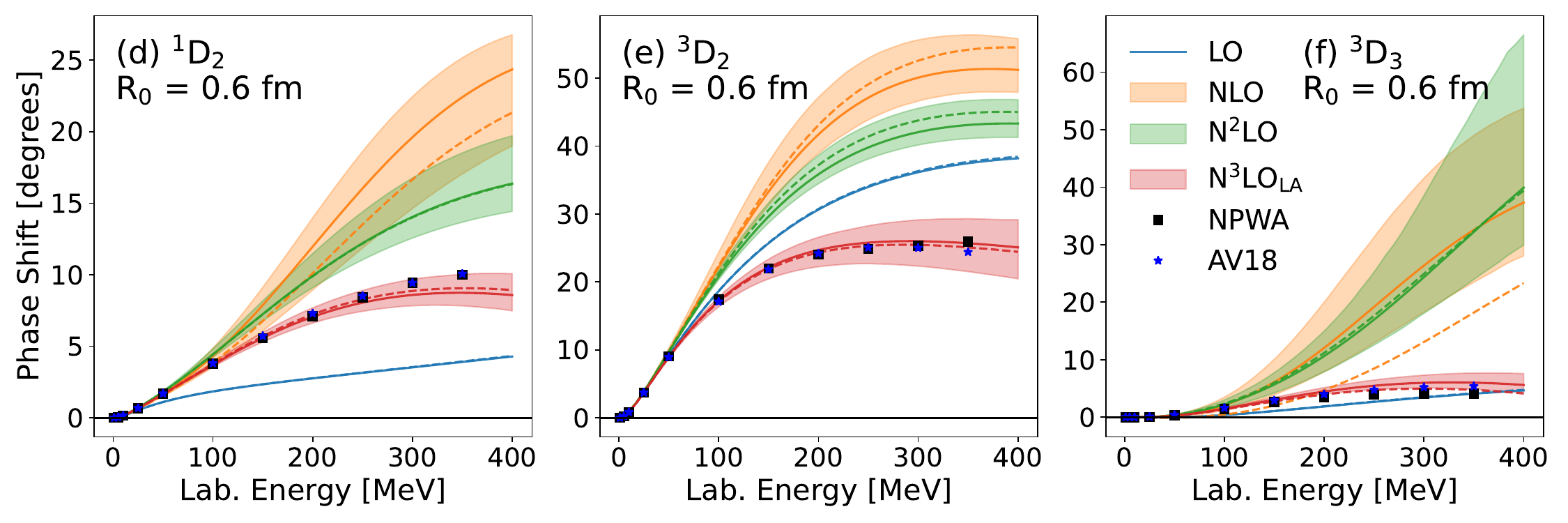}
    \includegraphics[scale=0.37]{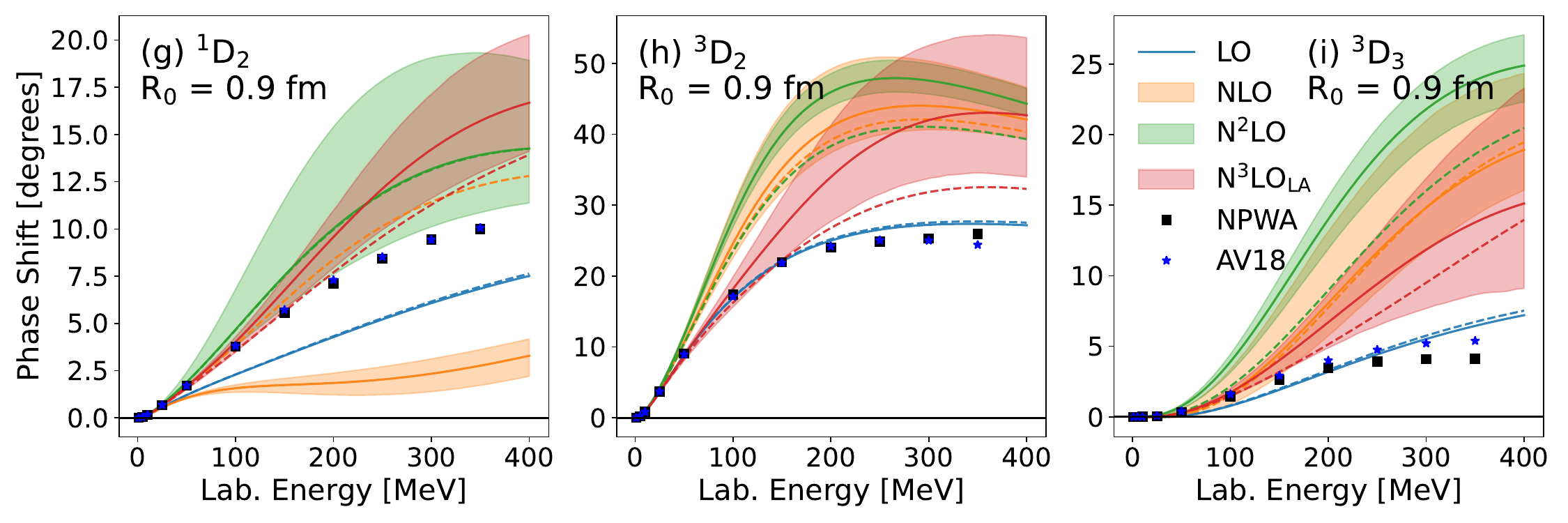}
    \caption{Same as Fig.~\ref{fig:ps_S}, but for the $^1D_2$, $^3D_2$ and $^3D_3$ partial waves.}
    \label{fig:ps_D}
\end{figure*}

The BUQEYE prescription Eq.~\eqref{eq:buqeye_L} holds under the assumption that the expansion coefficients $c_n$ are Gaussian distributed. 
The systematic uncertainties induced by truncating the EFT expansion does not necessarily need to follow such a distribution. 
Therefore, in this work, we turn to the EKM prescription~\eqref{eq:EKM} that makes fewer assumptions regarding the expansion coefficients $c_n$, i.e., it assumes a uniform distribution of these parameters. 
As mentioned earlier the EKM expression~\eqref{eq:EKM} provides only an estimate of the EFT truncation uncertainty and does not have a robust probabilistic interpretation, but see Ref.~\cite{Furnstahl:2015rha} for a mapping of the EKM error bars to the BUQEYE statistical distributions.
Therefore, as an ansatz, we choose $\Delta X^{k}_{\text{EKM}}$ to set the scale of our likelihood function, i.e., we take $\sigma_\text{theo,EKM} = \alpha \Delta X^{k}_{\text{EKM}}$, where $\alpha$ is a constant. 
We also add experimental uncertainties in quadrature to the theoretical truncation uncertainties, completing our ansatz for the variance of our likelihood function,
\begin{equation}
    \sigma^2 = \sigma_\text{exp}^2 + \sigma_\text{theo,EKM}^2\,.
    \label{eq:var}
\end{equation}
We emphasize that Eq.~\eqref{eq:var} is merely a model for the variance of our likelihood function, i.e., we do not claim to have derived it and no assumptions have been made regarding the distribution followed by the expansion coefficients $c_n$. 
We now turn to the form of the likelihood function. 
For this, we use the principle of maximum entropy which states that, among all real-valued functions with a specific variance $ \sigma^2$, the function that maximizes the differential entropy is a normal distribution with that variance. 
This fixes our likelihood function to be,
\begin{equation}
    \mathcal{L} \propto \prod_{i} \exp \bigg\{ -\frac{1}{2} \bigg( \frac{X^{\text{exp}}_i  -X^{\text{theo}}_i}{\sigma_i} \bigg)^2 \bigg\} \,,
    \label{eq:L}
\end{equation}
where $\sigma_i$ is given by Eq.~\eqref{eq:var}.

As mentioned above, in this work we take the EFT truncation uncertainty to be $\sigma_\text{i,theo,EKM} = \alpha \Delta X^{k}_{\text{EKM}}$. 
In the limit $\sigma_\text{i,theo}^2~\gg~\sigma_\text{i,exp}^2$ (which we have verified for all $i$), the parameter $\alpha$ turns out to be only an overall constant in the log-likelihood function. 
In the case of uninformative uniform priors on the LECs, the choice of $\alpha$, therefore, has negligible impact in our Bayesian analysis, eliminating the impact of differences of the EKM confidence interval at different orders. 
For simplicity, we take $\alpha=1$ and we have checked that changing it to $\alpha=1/2$ has no impact on our results. 
The experimental uncertainties $\sigma_\text{i,exp}$ are taken to be the uncertainties provided by the NPWA~\cite{Stoks:1993tb}.

The likelihood employed in this work, Eq.~\eqref{eq:L}, and that in Eq.~\eqref{eq:buqeye_L} are similar, with the only difference being the contribution of the EFT truncation uncertainty to the variance of the Gaussian likelihood, i.e., $\sigma_\text{theo,BUQEYE}$ vs $\sigma_\text{theo,EKM}$. 
In Appendix~\ref{sec:Buqeye_vs_EKM}, we compare these two quantities and show that these two choices are almost identical, leading to a similar Bayesian fit of our interactions. 
With this justification, we will henceforth focus on the posterior defined using Eq.~\eqref{eq:L} and leave a more detailed comparison with the BUQEYE approach for future work. 
Finally, we note that our approach to modeling theoretical truncation uncertainties does not incorporate the correlation of these uncertainties across different kinematic points~\cite{Wesolowski:2018lzj,Melendez:2019izc}. In Ref.~\cite{Svensson:2023twt}, it has been shown that incorporating such correlations doubles the variance of the marginal posteriors of all LECs, but it leaves the structure of the LEC posteriors unchanged. 
Here, we thus limit ourselves to the uncorrelated case, and will perform a more sophisticated analysis that includes these correlations in the future. 

For the fits at \NNLO and \NNNLO, we set $X^\mathrm{LO} = 0$ because we have otherwise found very poor Bayesian fits at high laboratory energies, with many samples including resonances or spurious bound states.
The reason for this is that, due to the large differences between the data and the LO predictions at high laboratory energies, the EKM uncertainty estimates are dominated by the poor LO predictions and are very large, removing the constraining power of any higher-energy data points and leading to spurious structures. Hence, we treat the LO contribution as an outlier in the expansion; see also Ref.~\cite{Svensson:2023twt} for a similar treatment of the LO predictions. 

Evaluating the EKM uncertainty, Eq.~\eqref{eq:EKM}, at a given order requires knowledge of the phase shifts at all lower orders.
Therefore, as an initial step to performing Bayesian fits, we first determine the LECs via a least-squares minimization of the objective function,
\begin{equation}
    \chi^2 = \frac{1}{m} \sum_i \bigg( \frac{X^{\text{exp}}_i  -X^{\text{theo}}_i}{\sigma_{i,\text{exp}}} \bigg)^2,
    \label{eq:chi2}
\end{equation}
where $m$ is the number of experimental data points included in the fit. 
At LO, the least-squares optimizations are done by fitting to the $^1S_0$ and $^3S_1$ partial waves up to a laboratory energy of $E_\text{max} = 50$ MeV. 
At NLO and N$^2$LO, we fit the $^1S_0$, $^3S_1$, $\epsilon_1$, $^1P_1$, $^3P_0$, $^3P_1$, and $^3P_2$ partial waves up to $E_\text{max} = 150$ MeV. 
For these orders, we fit to phase shift values at energies specified in Ref.~\cite{Gezerlis:2014zia}.
At N$^3$LO, we additionally include $^3D_1$, $\epsilon_2$, $^1D_2$, $^3D_2$, and $^3D_3$ in the fit and we fit up to $E_\text{max} = 250$ MeV, additionally including points at $200$ and $250$ MeV.\footnote{For the least-squares fit at $R_0 = 0.7$~fm alone, we chose $E_\text{max} = 350$ MeV in order to remove a spurious resonance in the $^1P_1$ channel at $E\approx 350$ MeV. 
We verified that this change of $E_\text{max}$ has a negligible effect on the phase shifts below 350 MeV in all other channels.}
The results of these fits serve as an order-by-order estimate of the EFT convergence that is used to estimate the EKM uncertainty Eq.~\eqref{eq:EKM} which, in turn, is used as an input for our Bayesian fits.
Note that the $\chi^2$ function used in the least-squares fit does not incorporate theoretical EFT truncation uncertainties. 
Therefore, the least-squares fits also serve as complementary analyses to the Bayesian fits and can be used to estimate the importance of modeling EFT truncation uncertainties when chiral EFT interactions are calibrated to scattering phase shifts. 

\begin{figure}[t]
    \centering
    \includegraphics[scale=0.53]{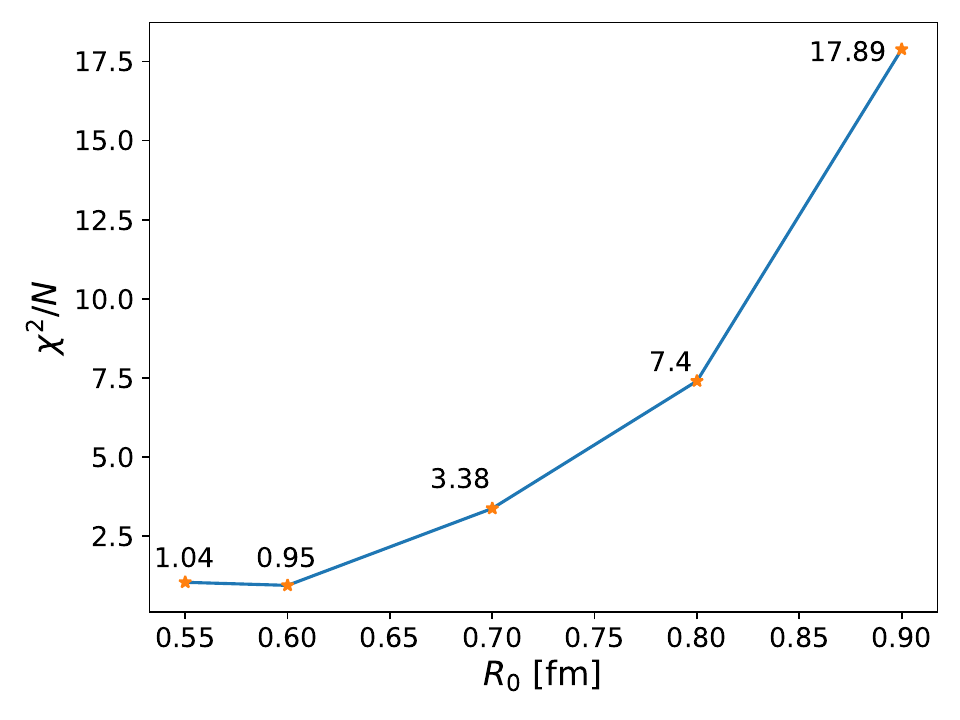}
    \caption{Objective function defined in Eq.~\eqref{eq:chi2} divided by the number of fit parameters $N$ (21 at \NNNLO) as a function of cutoff for our N$^3$LO$_{\rm LA}$ interactions. The calculated values are written explicitly in the figure.\\
    }
    \label{fig:chi2}
\end{figure}

In Fig.~\ref{fig:pot}, the left panel shows the local component of our N$^3$LO interactions in the $^1S_0$ channel, with the LECs determined via the above mentioned least-squares fits. 
The interaction becomes increasingly hard for smaller $R_0$ because the LO $^1S_0$ spectral LEC $\Tilde{C}_{1S0}$ increases rapidly with decreasing $R_0$; see right panel of Fig.~\ref{fig:pot}. 
This growth is required to counter the attractive pion-exchange potential that becomes increasingly singular for lower coordinate-space cutoffs~\cite{Huth:2017wzw}.

The Bayesian analyses are carried out by a Markov-chain Monte Carlo (MCMC) sampling of the posterior distribution defined in Eq.~\eqref{eq:bayes}. 
The details of the Bayesian fits, i.e., the partial waves involved and $E_\text{max}$ at a given order, are the same as in the least-squares fit. 
The MCMC sampling results in a large number of samples drawn from the posterior distribution. 
We used the \textsc{emcee} Python package~\cite{foreman2013emcee} to carry out the MCMC sampling and we used $5000$ walkers. 
To check the convergence of the MCMC sampling, i.e., to ensure that sufficient iterations were performed, we imposed that the total number of sampler iterations was larger than $50$ times the autocorrelation time for all the sampling parameters, i.e., for all the LECs. 
We also checked that our estimate of the autocorrelation time, as computed by the \textsc{emcee} package, remained stable (to within $1\%$) as a function of the sampler iteration. 
Once the MCMC sampling was terminated based on these convergence criteria, we discarded the first $2 \tau_\text{max}$ samples, where $\tau_\text{max}$ is the largest autocorrelation time across all sampling parameters, to make sure that we use only the ``burnt-in'' samples. 
We also thinned our sample chains by $0.5 \tau_\text{min}$ to remove any spurious correlations among our samples.
Finally, we found that a small percentage ($< 5\%$) of the samples at N$^3$LO contained spurious resonances. 
We discarded these samples and verified that this did not modify the posterior distribution significantly. 
In this manner, we obtained a posterior distribution over the entire parameter space, which at \NNNLO is a parameter space spanned by the 21 LECs. 
This posterior can be converted to posteriors over any two-body observable, such as $np$ phase shifts. 

\begin{figure}[t]
    \centering
    \includegraphics[scale=0.53]{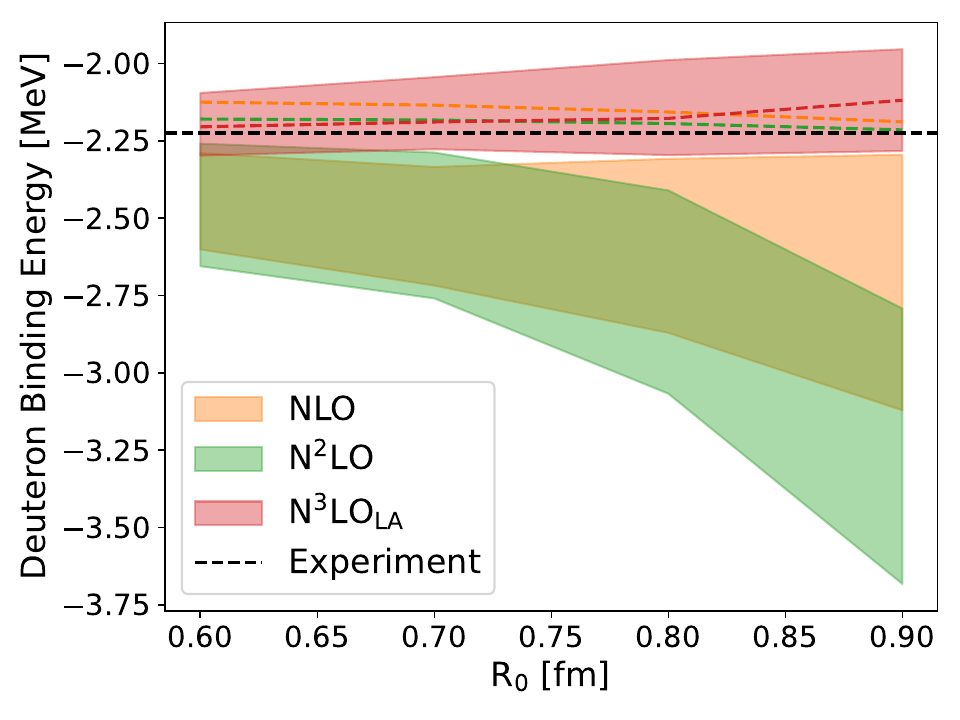}
    \caption{Deuteron binding energy as a function of cutoff for different orders, as indicated in the legend. 
    The bands correspond to the 95\% CL obtained from the Bayesian analysis whereas the dashed lines represent the least-squares results.
    The dashed black line shows the experimental binding energy.}
    \label{fig:deutron}
\end{figure}

In Fig.~\ref{fig:ps_S}, we show our results for the $np$ phase shifts in the $^1S_0$ and $^3S_1$-$^3D_1$ partial waves for both the least-squares fit (dashed lines) and the Bayesian fits (solid lines that represent the maximum of the posterior and the bands that correspond to the $95\%$ confidence level (CL) given the definition of uncertainty discussed above). 
In the top row, we demonstrate the effect of varying the cutoff $R_0$ at \NNNLO, i.e., we show results for our interactions N$^3$LO$_{\rm LA}$-09 to N$^3$LO$_{\rm LA}$-06.
In the middle (bottom) row, we vary the chiral EFT order with the cutoff fixed at $R_0 = 0.6$~fm ($R_0 = 0.9$~fm). 
Figs.~\ref{fig:ps_P} and~\ref{fig:ps_D} show similar results but for $P$ and $D$ waves respectively.
In general, regardless of the cutoff, we find good convergence order-by-order, with our results approaching the NPWA data with reduced uncertainty bands when going to higher orders. 
By comparing results for different cutoffs, we see that the softer interactions deviate significantly from the data at higher laboratory energies in several partial waves, especially the $D$ waves. 

\begin{figure*}
    \centering
    \includegraphics[scale=0.5]{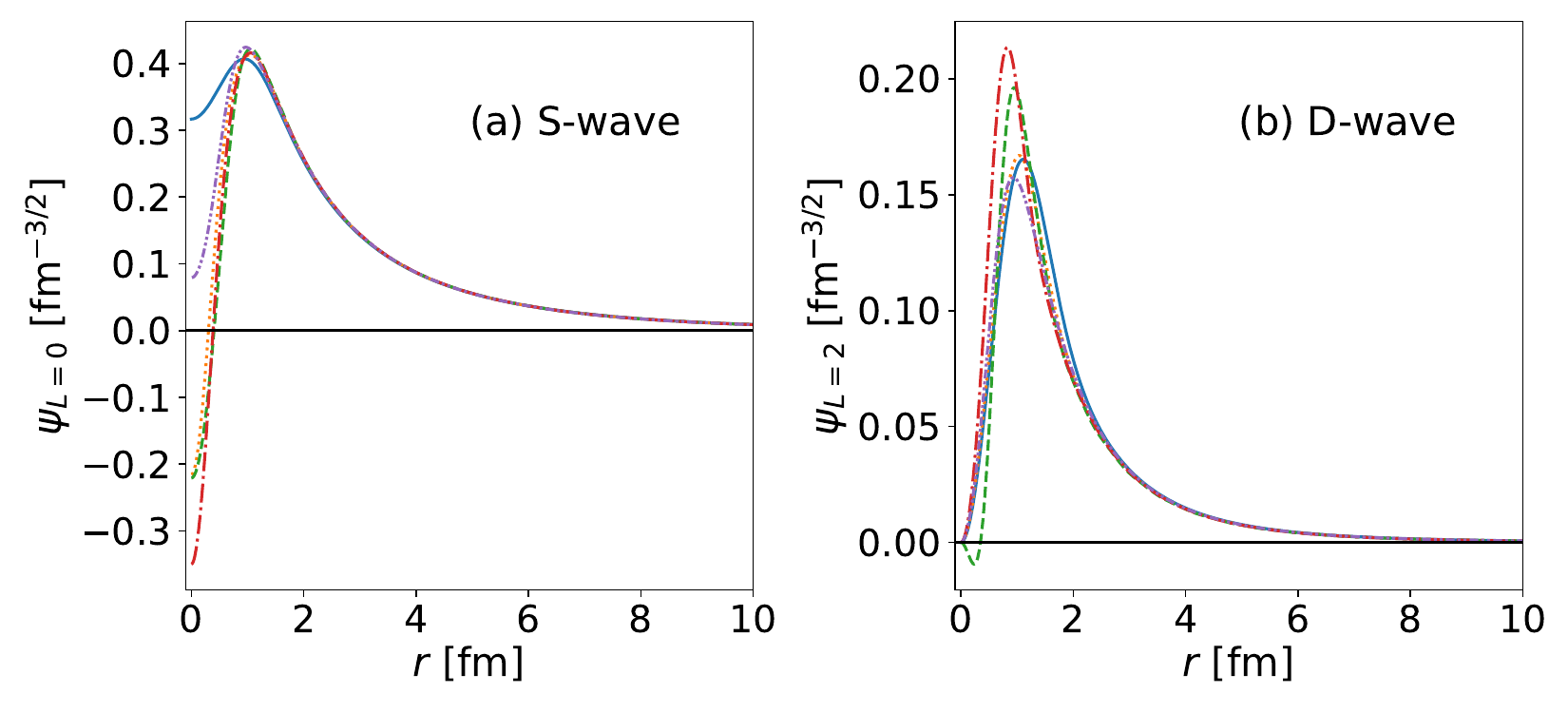}
    \includegraphics[scale=0.5]{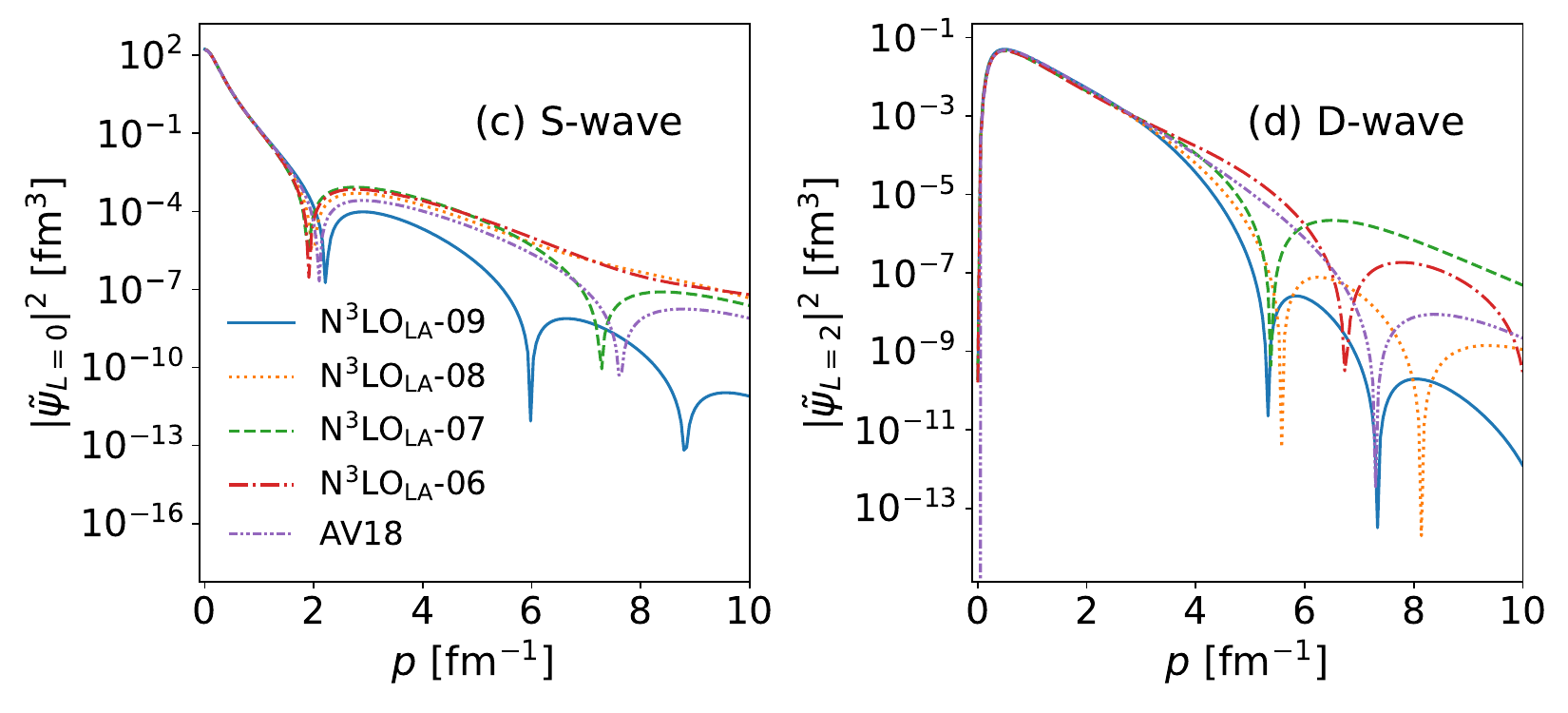}
    \caption{Deuteron wave function in coordinate space [panels (a) and (b)] and momentum space [panels (c) and (d)]. The results for different $R_0$ correspond to the least-squares phase shifts analyses. The AV18~\cite{Wiringa:1994wb} analysis is shown as a reference.}
    \label{fig:deutron_wf}
\end{figure*}

Several comments are in order.
First, note the large deviations of the LO phase shifts from the data in almost all channels.
This effect enhances the contribution of the LO phase shifts to the EKM uncertainties, which is why it is important to set $X^\mathrm{LO} = 0$ in Eq.~\eqref{eq:EKM} for the fits at N$^2$LO and N$^3$LO.
Second, it is important to keep in mind that the uncertainty bands shown here, reflecting the Bayesian posteriors obtained in the fit, describe the LEC uncertainties and are not necessarily the same as the bands obtained following the EKM prescription {\it a posteriori}, as was done in Ref.~\cite{Epelbaum:2014efa}. 
This can be understood by considering an LO interaction where the EKM uncertainties widen with the laboratory energy due to the almost constant phase shift.
However, parts of such an EKM uncertainty band would not be accessible in an actual fit.
For example, describing phase shifts that decrease with the laboratory energy, as provided by the lower bound of an EKM uncertainty band, requires effective-range contributions which are only provided at NLO.
Therefore, the results of a Bayesian fit will differ and show a less dramatic energy dependence. For this reason, we do not show the 95\% CL bands for the LO results, and we plot only the least-squares and the maximum posterior results (dashed and solid lines, respectively). 
Finally, our interactions at lower orders are not fit to the $D$ waves but the \NNNLO interactions are.
As a consequence, the differences between \NNLO and \NNNLO bands in the D waves are larger than the differences between \NLO and \NNLO.

The impact of the inclusion of the EFT truncation uncertainties that guide the Bayesian fits can be gauged by comparing the solid and dashed lines in all panels in Figs.~\ref{fig:ps_S}-\ref{fig:ps_D}.
Especially at lower cutoffs, the impact of the EFT truncation uncertainties is important and the Bayesian analysis yields a better reproduction of the data at lower energies as compared to the least-squares fits. At higher energies, the least-squares fits perform better generally. Note however that, since we do a global fit involving several partial waves, there are some channels in which the Bayesian fit gives better results at higher energies, e.g., for the mixing angle $\epsilon_1$ for our N$^3$LO$_{\rm LA}$ interactions.

In general, we also find that the high-cutoff interactions give a better reproduction of the NPWA data than low-cutoff interactions, see the top rows in Figs.~\ref{fig:ps_S},~\ref{fig:ps_P} and~\ref{fig:ps_D} where the phase shifts at \NNNLO are compared for different cutoffs.
The Bayesian fits clearly demonstrate the reduction of uncertainty when decreasing $R_0$, which is equivalent to increasing the momentum-space cutoff $\Lambda_c$. 
In order to stress this point, in Fig.~\ref{fig:chi2} we show the objective function defined in Eq.~\eqref{eq:chi2} divided by the number of fit parameters (21 at \NNNLO) as a function of $R_0$. 
We see the strong decrease in $\chi^2$ as interactions become harder, before plateauing and eventually increasing at $R_0 \approx 0.55$~fm.

The LECs that we obtain for our N$^3$LO$_{\rm LA}$ interactions are given in Table~\ref{tab:LEC_N3LO} for both fitting strategies: least-squares fits and Bayesian fits (maximum posterior estimate). Similar results for the LECs at LO, \NLO, and \NNLO are given in Appendix~\ref{sec:LECs_lower_orders}.

\begin{table*}[t]
\centering
\tabcolsep=0.5cm
\def\arraystretch{1.5}
\begin{tabular}{c|cccc|cccc}
\hline
LEC & \multicolumn{4}{c|}{Maximum posterior estimate} & \multicolumn{4}{c}{Least-squares fit}  \\
\hline
                & $0.9$~fm & $0.8$~fm & $0.7$~fm & $0.6$~fm & $0.9$~fm & $0.8$~fm & $0.7$~fm & $0.6$~fm\\
\hline
$C_S$~[fm$^2$]       & 2.371 & 4.784 & 13.293 & 27.649 & 3.698 & 5.436 & 15.28 & 27.595 \\
$C_T$~[fm$^2$]      & 0.785 & 0.79 & 2.503 & 2.304 & 1.029 & 0.704 & 2.552 & 2.742 \\
$C_1$~[fm$^4$]     & $-0.098$ & $-0.001$ & 0.276 & 0.365 & 0.167 & 0.088 & 0.416 & 0.352 \\
$C_2$~[fm$^4$]      & 0.129 & 0.011 & 0.032 & 0.013 & 0.086 & 0.003 & 0.023 & 0.016 \\
$C_3$~[fm$^4$]      & 0.031 & 0.009 & 0.075 & 0.003 & 0.013 & $-0.003$ & 0.077 & 0.004 \\
$C_4$~[fm$^4$]      & $-0.0$ & $-0.019$ & 0.011 & 0.03 & 0.025 & $-0.019$ & 0.004 & 0.031 \\
$C_5$~[fm$^4$]      & $-1.765$ & $-2.01$ & $-2.358$ & $-2.231$ & $-2.047$ & $-2.072$ & $-2.309$ & $-2.168$ \\
$C_6$~[fm$^4$]      & 0.043 & 0.07 & 0.154 & 0.241 & 0.118 & 0.074 & 0.16 & 0.291 \\
$C_7$~[fm$^4$]      & $-0.171$ & $-0.15$ & $-0.175$ & $-0.186$ & $-0.217$ & $-0.144$ & $-0.165$ & $-0.179$ \\
$D_1$~[fm$^6$]      & $-0.013$ & 0.007 & 0.019 & 0.023 & $-0.009$ & 0.007 & 0.029 & 0.023\\
$D_2$~[fm$^6$]       & 0.01 & 0.006 & 0.007 & 0.008 & 0.025 & 0.007 & 0.01 & 0.008 \\
$D_3$~[fm$^6$]      & $-0.012$ & $-0.012$ & $-0.006$ & $-0.004$ & $-0.02$ & $-0.017$ & $-0.006$ & $-0.005$ \\
$D_4$~[fm$^6$]       & $-0.004$ & 0.0 & 0.007 & 0.008 & 0.009 & 0.007 & 0.012 & 0.009 \\
$D_5$~[fm$^6$]      & 0.111 & 0.032 & $-0.01$ & $-0.053$ & 0.148 & 0.076 & 0.009 & $-0.058$ \\
$D_6$~[fm$^6$]       & 0.042 & 0.035 & 0.026 & 0.01 & $-0.037$ & $-0.01$ & 0.001 & 0.001 \\
$D_7$~[fm$^6$]      & 0.042 & 0.033 & 0.031 & 0.024 & 0.07 & 0.036 & 0.037 & 0.029 \\
$D_8$~[fm$^6$]      & $-0.051$ & $-0.03$ & $-0.028$ & $-0.019$ & $-0.069$ & $-0.036$ & $-0.032$ & $-0.02$ \\
$\tilde{D}_{11}$~[fm$^6$]      & 0.005 & $-0.09$ & $-0.101$ & $-0.092$ & $-0.163$ & $-0.133$ & $-0.12$ & $-0.092$ \\
$\tilde{D}_{12}$~[fm$^6$]     & $-0.059$ & $-0.009$ & $-0.007$ & $-0.004$ & $-0.064$ & $-0.007$ & $-0.005$ & $-0.006$ \\
$\tilde{D}_{13}$~[fm$^6$]     & $-0.033$ & $-0.042$ & $-0.036$ & $-0.032$ & $-0.083$ & $-0.044$ & $-0.035$ & $-0.037$ \\
$\tilde{D}_{15}$~[fm$^6$]     & $-0.001$ & $-0.008$ & 0.005 & 0.015 & 0.061 & 0.037 & 0.023 & 0.016 \\
\hline
\end{tabular}
\caption{LECs for our N$^3$LO$_{\rm LA}$ interactions for different cutoffs. We give results obtained from the Bayesian analyses (left) and least-squares fits (right).
For the former, the quoted LECs correspond to the maximum of the posterior distribution.}
\label{tab:LEC_N3LO}
\end{table*}

\section{Deuteron}
\label{sec:deuteron}

Finally, we study properties of the deuteron, which can be completely described by the $NN$ interactions developed here.
In this work, we do not fit the interactions to reproduce any properties of the deuteron, and instead choose experimental data on deuteron properties as a benchmark for our interactions. 

In Fig.~\ref{fig:deutron}, we show the deuteron binding energy as a function of cutoff, for different orders. The bands represent the Bayesian posteriors at the 95\% CL whereas the least-squares results are shown as dashed lines. 
We see that, at \NNNLO, the least-squares results which do not include any theoretical uncertainty estimates are compatible with the predictions obtained from the Bayesian fits. 
However, for \NLO and \NNLO, there is a clear difference between the two analyses. This, once again, illustrates the importance of incorporating theoretical uncertainties when EFTs are calibrated to experimental data.
Also, note that the uncertainties decrease significantly with the chiral order as well as when decreasing $R_0$, which is consistent with our results for the phase shifts. 
Our N$^3$LO$_\text{LA}$ interactions reproduce the experimental deuteron binding energy within theoretical uncertainties for all considered cutoffs.

In Fig.~\ref{fig:deutron_wf}, we show results for the deuteron wave function. 
In the top row, we show the coordinate space wavefunction in the $S$ wave, $\psi_{L=0}(r)$, and $D$ wave, $\psi_{L=2}(r)$. 
Note that these wavefunctions are related to their components $u(r)$ and $w(r)$, which are sometimes reported in the literature~\cite{Epelbaum:2004fk,Reinert:2017usi,Piarulli:2014bda}, as $\psi_{L=0}(r) = u(r)/r$ and $\psi_{L=2}(r) = w(r)/r$. 
In the bottom row, we show the momentum-space representation of the wavefunctions in the $S$ wave, $\tilde{\psi}_{L=0}(p)$, and $D$ wave, $\tilde{\psi}_{L=2}(p)$, showing the squared momentum-space wavefunctions in a logarithmic scale. 
As our fitting code is written in momentum space, we first solve for the momentum-space wavefunctions and then obtain the coordinate-space representations as,
\begin{align}
    \psi_{L=0}(r) &= \sqrt{ \frac{2}{\pi} } \int_0^\infty dp p^2 j_0(pr) \tilde{\psi}_{L=0}(p) \,, \\
    \psi_{L=2}(r) &=\sqrt{ \frac{2}{\pi} } \int_0^\infty dp p^2 j_2(pr) \tilde{\psi}_{L=2}(p)\,.
\end{align}
In coordinate space, our models do not contain any oscillations in the wave function at intermediate or large $r$, unlike some chiral interactions developed in the past~\cite{Epelbaum:2004fk,Entem:2003ft}. 
Interactions at different cutoffs are identical at large $r$ but have different small-$r$ behavior.
In the $D$ wave, the harder interactions, i.e., N$^3$LO$_{\rm LA}$-06 and N$^3$LO$_{\rm LA}$-07, lead to more pronounced peaks as compared to their softer counterparts N$^3$LO$_{\rm LA}$-08 and N$^3$LO$_{\rm LA}$-09. 

We have used the deuteron wave functions at \NNNLO to calculate other deuteron observables, which we show in Table~\ref{tab:deuteron}. 
We give results for both our Bayesian fits (at the 95\% CL) and least-squares fits. 
For the Bayesian fits, our results are in good agreement with the experimental data within theoretical uncertainties, except for the quadrupole moment which is to be expected since we do not take electromagnetic two-body currents into account.
Note that the agreement between the median model predictions and the experiment is better for interactions with smaller coordinate space cutoffs. 
Furthermore, these harder interactions have smaller uncertainties, similar to what was observed in the phase shift analyses. 
The least-squares results are also in good agreement with the experimental data, and the agreement improves for harder interactions. 
We note that, since we fit our interactions to the phase shifts in the $^3S_1$-$^3D_1$ channel, it is not surprising that we get reasonable deuteron properties. However, this calculation still serves as a valuable model check. 
Finally, note that our interactions result in $D$-state probabilities $P_d \gtrapprox 6\%$, which is larger that what has been found for other local \NNNLO interactions~\cite{Piarulli:2014bda,Saha:2022oep}.

\begin{table*}[t]
\centering
\tabcolsep=0.22cm
\def\arraystretch{1.5}
\begin{tabular}{c|cccc|cccc|c}
\hline
Observable & \multicolumn{4}{c|}{Bayesian fit} & \multicolumn{4}{c|}{Least-squares fit} & \multicolumn{1}{c}{Experiment}  \\
\hline
                & $0.9$~fm & $0.8$~fm & $0.7$~fm & $0.6$~fm & $0.9$~fm & $0.8$~fm & $0.7$~fm & $0.6$~fm\\
\hline
$E_d$ [MeV] & $-2.1^{+0.2}_{-0.2}$ & $-2.1^{+0.1}_{-0.2}$ &  $-2.2^{+0.1}_{-0.1}$&  $-2.2^{+0.1}_{-0.1}$ & $-2.119$  & $-2.177$ & $-2.189$ & $-2.205$ & $-2.224575(9)$ \\
$Q_d$ [fm$^2$]  & $0.270^{+0.005}_{-0.005}$ & $0.268^{+0.004}_{-0.005}$ & $0.268^{+0.004}_{-0.004}$ &  $0.268^{+0.003}_{-0.003}$ & 0.278  & 0.269 & 0.265 & 0.264 & 0.2859(3) \\
$\eta_d$  & $0.024^{+0.002}_{-0.002}$ & $0.024^{+0.002}_{-0.002}$ &  $0.024^{+0.001}_{-0.001}$ & $0.025^{+0.001}_{-0.001}$  & 0.0247  & 0.0244 & 0.0241 & 0.0233 & 0.0256(4) \\
$\sqrt{\langle r^2 \rangle ^d_m}$ [fm] & $1.97^{+0.06}_{-0.06}$ & $1.99^{+0.05}_{-0.05}$ &  $1.98^{+0.04}_{-0.04}$ & $1.97^{+0.03}_{-0.03}$  & 1.983  & 1.975 & 1.973 & 1.97 & 1.9753(11)\\
$A_S$ [fm$^{-1/2}$] & $0.85^{+0.03}_{-0.03}$ & $0.87^{+0.03}_{-0.02}$ & $0.87^{+0.02}_{-0.02}$ &  $0.88^{+0.02}_{-0.02}$ & 0.860  & 0.875 & 0.878 & 0.882 & 0.8846(9) \\
$P_d$ [\%]& $6.5^{+0.8}_{-0.8}$ & $6.0^{+0.7}_{-0.6}$ & $6.3^{+0.5}_{-0.5}$ &  $6.1^{+0.4}_{-0.4}$ & 6.64  & 6.00 & 6.16 & 6.26 & -- \\
\hline
\end{tabular}
\caption{Properties of the deuteron for our N$^3$LO$_{\rm LA}$ interactions, for different cutoffs. For the Bayesian fit, the error bars are quoted at the 95\% CL. Here, $E_d$ is the binding energy, $Q_d$ is the quadrupole moment, $\eta_d$ is the asymptotic D/S ratio, $\sqrt{\langle r^2 \rangle ^d_m}$ is the root-mean-square matter radius, $A_S$ is the asymptotic $S$-wave normalization and $P_d$ is the $D$-state probability. The experimental value for $E_d$ is from Ref.~\cite{VanDerLeun:1982bhg}, $Q_d$ is from Ref.~\cite{Bishop:1979zz}, $\eta_d$ is from Ref.~\cite{Rodning:1990zz}, $\sqrt{\langle r^2 \rangle ^d_m}$ is from Ref.~\cite{Friar:1997js} and $A_S$ is from Ref.~\cite{Ericson:1982ei}.}
\label{tab:deuteron}
\end{table*}

\section{Summary and Outlook}
\label{sec:conclusion}

We have constructed maximally local interactions at \NNNLO in $\Delta$-less chiral EFT. 
Our interactions include a total of 21 contact operators at \NNNLO, out of which four are nonlocal. 
We have also included all pion-exchange contributions consistently up to \NNNLO. 
We have employed the method of Bayesian statistics in order to calibrate our maximally local interactions to $np$ phase shifts, explicitly taking into account EFT truncation uncertainties.
We have demonstrated the importance of incorporating these uncertainties in the fit. 
Finally, we have explored a range of cutoffs that is significantly larger than what is typically explored in the literature, showing explicitly that high-cutoff interactions lead to smaller EFT truncation errors and a better reproduction of experimental data. 
Furthermore, it is expected that these interactions significantly reduce uncertainties due to the violation of the Fierz rearrangement freedom, particularly from $3N$ interactions~\cite{Lynn:2015jua}.
We will demonstrate this explicitly in a forthcoming work.   

Future QMC calculations using the interactions developed in this work will allow us to reduce theoretical uncertainties for nuclei and dense nuclear matter, which is important to improve nuclear-physics inputs to nuclear structure and reactions, and for analyses of multi-messenger observations of neutron stars and related simulations. 

\acknowledgements

We thank Joe Carlson, Andreas Ekstr{\"o}m, Richard Furnstahl, Stefano Gandolfi, Ruprecht Machleidt, Joshua Martin, Jordan Melendez, Sam Novario, Daniel Phillips, and Ronen Weiss for useful discussions. 
R.S. acknowledges support from the Nuclear Physics from Multi-Messenger Mergers (NP3M) Focused Research Hub which is funded by the National Science Foundation under Grant No. 21-16686, and by the Laboratory Directed Research and Development program of Los Alamos National Laboratory under Project No. 20220541ECR.
The work of J.E.L., L.H., and A.S. was supported by the European Research Council (ERC) Grant No.~307986 STRONGINT and under the European Union's Horizon 2020 research and innovation programme (Grant Agreement No.~101020842).
I.T. was supported by the U.S. Department of Energy, Office of Science, Office of Nuclear Physics, under contract No.~DE-AC52-06NA25396, by the Laboratory Directed Research and Development program of Los Alamos National Laboratory under Project No. 20220541ECR, and by the U.S. Department of Energy, Office of Science, Office of Advanced Scientific Computing Research, Scientific Discovery through Advanced Computing (SciDAC) NUCLEI program.

\appendix

\begin{widetext}

\section{Fourier transformations of \NNNLO contact operators}
\label{sec:FTs}

In this appendix, we give the Fourier transform (FT) of the \NNNLO contact operators for a local regulator $f(r)$; see Sec.~\ref{sec:regulators} for the specific form of $f(r)$ used in this work. We start with the local operators $D_1$ -- $D_4$ of Eq.~\eqref{eq:n3lo}. Let us denote by $V_{\tau,\sigma}$ the part of the operator that contains the spin-isospin operator structures (momentum-independent part) as well as the LEC. We find
\begin{align}
    \bra{\fet r} V_{\tau,\sigma} q^4   \ket{\psi} = V_{\tau,\sigma} \Delta^2 f(r) \psi(\fet r).
\end{align}
For typical regulators of the form $f(r)=a \exp(-(r/R_0)^n)$, the squared Laplacian is given as
\begin{align}
\Delta^2 f(r)&= n
\left[-\frac{(n+1)(n-1)(n-2)r^{n-4}}{R_0^n}+\frac{n(7n+1)(n-1)r^{2n-4}}{R_0^{2n}}\right. \nonumber\\
&\quad-\left. \frac{2n^2(3n-1)r^{3n-4}}{R_0^{3n}}+\frac{n^3 r^{4n-4}}{R_0^{4n}}
\right]a \exp(-(r/R_0)^n).
\end{align}

Next, the FT for the N$^3$LO spin-orbit term $V_{\tau} i\fet q^2  (\fet \sigma_1+ \fet\sigma_2)\cdot(\fet q \times \fet k)$, i.e., operators $D_5$ and $D_6$ of Eq.~\eqref{eq:n3lo}, can be written as 
\begin{align}
    \bra{\fet r} iV_{\tau}  \fet q^2  \fet L \cdot \fet S \ket{\psi} &=V_{\tau} \left(\frac{\partial_r^3 f(r)}{r}+ \frac{2\partial_r^2 f(r)}{r^2}-\frac{2\partial_r f(r)}{r^3}\right) \fet L \cdot \fet S \psi(\fet r)\,,
\end{align}
where we have set $\fet S = \fet \sigma_1+ \fet\sigma_2$. Next, we consider the FT of the tensor operators of the form $V_{\tau}\, 
\fet q^2 \fet \sigma_1 \cdot \fet q \, \fet \sigma_2 \cdot \fet q $. This corresponds to operators $D_7$ and $D_8$ in Eq.~\eqref{eq:n3lo}. The result is given by
\begin{align}
    \bra{\fet r}V_{\tau}\, 
\fet q^2 \fet \sigma_1 \cdot \fet q \, \fet \sigma_2 \cdot \fet q  \ket{\psi} &= V_{\tau} 
\left(\fet \sigma_1 \cdot \fet \sigma_2 \, \left(\frac{\partial_r^3 f(r)}{r}+\frac{2\partial_r^2 f(r)}{r^2}-\frac{2\partial_r f(r)}{r^3}\right) \right. \nonumber \\
&\left. \quad + 3\, \fet\sigma_1 \cdot \hat{\fet r} \, \fet \sigma_2 \cdot \hat{\fet r} \left(\frac{ \partial_r^4 f(r)}{3}+ \frac{\partial_r^3 f(r)}{3 r} - \frac{2\partial_r^2 f(r)}{r^2} + \frac{2\partial_r f(r)}{r^3} \right)\right) \psi(\fet r) \,.  
\label{eq:FT_q2_qten}
\end{align}

We now turn to the nonlocal operators used in this work, quoted in Eq.~\eqref{eq:nonlocal_ops}. First, let us consider the operators proportional to $\tilde{D}_{11}$ and $\tilde{D}_{12}$. These are of the form $V_{\sigma \tau} (\fet q \times \fet k)^2$. We find
\begin{align}
    \bra{\fet r} V_{\sigma \tau} \fet L^2\ket{\psi} &= - V_{\sigma \tau}\left(\frac{\partial_r^2 f(r)}{r^2}- \frac{\partial_r f(r)}{r^3} \right) \fet L^2+ 2 V_{\sigma \tau} \left( \frac{\partial_r f(r)}{r} \Delta + \left(\frac{\partial_r^2 f(r)}{r^2}- \frac{\partial_r f(r)}{r^3} \right) \fet r\cdot \nabla  \right) \psi(\fet r)  \,.
\end{align}

Next, consider the operator proportional to $\tilde{D}_{13}$. For this, we find
\begin{align}
    \bra{\fet r} \fet k^2 \fet \sigma_1\cdot \fet q \fet\sigma_2 \cdot \fet q  \ket{\psi} &= \frac{1}{4} \bra{\fet r}  \fet q^2 \fet \sigma_1 \cdot \fet q \, \fet \sigma_2 \cdot \fet q  \ket{\psi} \nonumber \\ &+ \left(-\frac{\partial_r f(r)}{r^2} + \frac{\partial_r^2 f(r)}{r}\right) \left( \fet \sigma_1 \cdot \hat{\fet r} \, \, \fet \sigma_2 \cdot \nabla + \fet \sigma_2 \cdot \hat{\fet r} \, \, \fet \sigma_1 \cdot \nabla + \fet \sigma_1 \cdot \fet \sigma_2  \hat{\fet r} \cdot \nabla \right) \psi(\fet r) \nonumber \\
    &+ \left( 3\frac{\partial_r f(r)}{r^2} -3 \frac{\partial_r^2 f(r)}{r} + \partial^3_r f(r) \right) \fet \sigma_1 \cdot \hat{\fet r} \, \, \fet \sigma_2 \cdot \hat{\fet r} \, \,\hat{\fet r} \cdot \nabla \psi(\fet r) \nonumber \\ &+ \left( \frac{\partial_r f(r)}{r} \left( \fet \sigma_1 \cdot \fet \sigma_2 - \fet \sigma_1 \cdot \hat{\fet r} \, \, \fet \sigma_2 \cdot \hat{\fet r} \right) + \fet \sigma_1 \cdot \hat{\fet r} \, \, \fet \sigma_2 \cdot \hat{\fet r} \, \, \partial^2_r f(r) \right) \Delta\psi(\fet r).
\end{align}
Note that the first term in the above result, i.e., $\bra{\fet r}  \fet q^2 \fet \sigma_1 \cdot \fet q \, \fet \sigma_2 \cdot \fet q  \ket{\psi}$, is given in Eq.~\eqref{eq:FT_q2_qten}.

The last operator we have to consider is $\tilde{D}_{15}$ of Eq.~\eqref{eq:nonlocal_ops} which has the form $(\fet \sigma_1 \cdot \fet L) (\fet \sigma_2 \cdot \fet L)$. Note that
\begin{align}
(\fet L \cdot \fet S)^2 
&=2 \left[ \fet q^2 \fet k^2- (\fet q \cdot \fet k)^2 +  (\fet \sigma_1 \cdot \fet L) (\fet \sigma_2 \cdot \fet L) \right]\nonumber \,.
\end{align}
Therefore, the coordinate space expression of $(\fet \sigma_1 \cdot \fet L) (\fet \sigma_2 \cdot \fet L)$ is equivalent to considering the FT of the three operators $(\fet L \cdot \fet S)^2$, $\fet q^2 \fet k^2$, and $(\fet q \cdot \fet k)^2$. For these three cases, the FT is given by
\begin{align}
\bra{\fet r} (\fet L \cdot \fet S)^2 \ket{\psi} 
&= -\left( \frac{\partial_r^2 f(r)}{r^2}-\frac{\partial_r f(r)}{r^3} \right) ( \fet L \cdot \fet S)^2 \psi(\fet r) \nonumber \\
&+ \frac{\partial_r f(r)}{r} (\fet S \times \nabla)^2 \psi(\fet r) + \left( \frac{\partial_r^2 f(r)}{r^2}-\frac{\partial_r f(r)}{r^3} \right) (\fet S \times \fet r)\cdot(\fet S\times \nabla)\psi(\fet r)\,,
\end{align}
\begin{align}
\bra{\fet r} \fet q^2 \fet k^2 \ket{\psi} &= \frac14 \left( \Delta^2 f(r)\right)\psi(\fet r)+ \partial_r^2 f(r)\Delta\psi(\fet r)+ \frac{\partial_r^3 f(r)}{r}\fet r \cdot \nabla \psi(\fet r) \nonumber \\
&\quad +2\left(\frac{\partial_r f(r)}{r}\Delta+\left(  \frac{\partial_r^2 f(r)}{r^2}- \frac{\partial_r f(r)}{r^3}
 \right)\fet r \cdot \nabla\right)\psi(\fet r),
\end{align}
\begin{align}
\bra{\fet r} (\fet q\cdot \fet k)^2 \ket{\psi} &= \frac14 \left( \Delta^2 f(r)\right)\psi(\fet r)+\frac{\partial_r^3 f(r)}{r} \fet r \cdot \nabla   \psi(\fet r) +\left(\frac{\partial_r^2 f(r)}{r^2}- \frac{\partial_r f(r)}{r^3}\right)\left(\fet r \cdot \nabla\right)^2\psi(\fet r) \nonumber \\
 &\quad +\left(\frac{\partial_r f(r)}{r}\Delta+\left(\frac{\partial_r^2 f(r)}{r^2}- \frac{\partial_r f(r)}{r^3} \right) \fet r \cdot \nabla \right)\psi(\fet r).
\end{align}

\end{widetext}

\section{Partial-wave decomposition of the contact interactions}
\label{sec:PWB}

The matrix elements of the \NNNLO contact operators, presented in Sec.~\ref{subsec:contacts}, in the partial-wave basis are given as follows
\begin{widetext}
\begin{align}
    \langle ^1S_0 | V^{(4)}_{\text{cont}} | ^1S_0 \rangle &= D^1_{1S0} p^2 p^{\prime 2} + D^2_{1S0} (p^4 + p^{\prime 4} ), \nonumber \\ 
    \langle ^3S_1 | V^{(4)}_{\text{cont}} | ^3S_1 \rangle &= D^1_{3S1} p^2 p^{\prime 2} + D^2_{3S1} (p^4 + p^{\prime 4} ),  \nonumber \\ 
    \langle ^1P_1 | V^{(4)}_{\text{cont}} | ^1P_1 \rangle &= D_{1P1} p p^{\prime} (p^2 + p^{\prime 2}), \nonumber \\ 
    \langle ^3P_1 | V^{(4)}_{\text{cont}} | ^3P_1 \rangle &= D_{3P1} p p^{\prime} (p^2 + p^{\prime 2}), \nonumber \\ 
    \langle ^3P_0 | V^{(4)}_{\text{cont}} | ^3P_0 \rangle &= D_{3P0} p p^{\prime} (p^2 + p^{\prime 2}), \nonumber \\ 
    \langle ^3P_2 | V^{(4)}_{\text{cont}} | ^3P_2 \rangle &= D_{3P2} p p^{\prime} (p^2 + p^{\prime 2}), \nonumber \\ 
    \langle ^1D_2 | V^{(4)}_{\text{cont}} | ^1D_2 \rangle &= D_{1D2} p^2 p^{\prime 2}, \nonumber \\ 
    \langle ^3D_2 | V^{(4)}_{\text{cont}} | ^3D_2 \rangle &= D_{3D2} p^2 p^{\prime 2}, \nonumber \\
    \langle ^3D_1 | V^{(4)}_{\text{cont}} | ^3D_1 \rangle &= D_{3D1} p^2 p^{\prime 2}, \nonumber \\ 
    \langle ^3D_3 | V^{(4)}_{\text{cont}} | ^3D_3 \rangle &= D_{3D3} p^2 p^{\prime 2}, \nonumber \\ 
    \langle ^3S_1 | V^{(4)}_{\text{cont}} | ^3D_1 \rangle &= D^1_{3D1-3S1} p^2 p^{\prime 2} + D^2_{3D1-3S1} p^4, \nonumber \\ 
    \langle ^3D_1 | V^{(4)}_{\text{cont}} | ^3S_1 \rangle &= D^1_{3D1-3S1} p^2 p^{\prime 2} + D^2_{3D1-3S1} p^{\prime 4}, \nonumber \\ 
    \langle ^3P_2 | V^{(4)}_{\text{cont}} | ^3F_2 \rangle &= D_{3F2-3P2} p^3 p^{\prime}, \nonumber \\ 
    \langle ^3F_2 | V^{(4)}_{\text{cont}} | ^3P_2 \rangle &= D_{3F2-3P2} p p^{\prime 3} ,
\end{align}
where we have used the spectroscopic LECs that are related to the operator LECS by
\begin{align}
    D^1_{1S0} &= \frac{2 \pi}{3} [ 20 D_1 + 20 D_2 - 60 D_3 -60 D_4 -20 D_7 -20 D_8 + 4 \tilde{D}_{11} + 4 \tilde{D}_{12} - \tilde{D}_{13} -4 \tilde{D}_{15}  ], \nonumber \\ 
    D^2_{1S0} &= \pi [ 4 D_1 + 4 D_2 - 12 D_3 -12 D_4 -4 D_7 -4 D_8  - \tilde{D}_{13} ], \nonumber \\ 
    D^1_{3S1} &= \frac{2 \pi}{9} [ 60 D_1 - 180 D_2 + 60 D_3 - 180 D_4 +20 D_7 -60 D_8 + 12 \tilde{D}_{11} - 36 \tilde{D}_{12} + \tilde{D}_{13} + 4 \tilde{D}_{15} ], \nonumber \\ 
    D^2_{3S1} &= \frac{\pi}{3} [ 12 D_1 - 36 D_2 + 12 D_3 - 36 D_4 +4 D_7 -12 D_8 + \tilde{D}_{13}], \nonumber \\ 
    D_{1P1} &= \frac{-16 \pi}{3} [  D_1 - 3 D_2 - 3 D_3 + 9 D_4 - D_7 +4 D_8 ], \nonumber \\ 
    D_{3P1} &= \frac{-2 \pi}{3} [ 8 D_1 + 8 D_2 +8  D_3 +8 D_4 - 4 D_5 - 4 D_6 + 12 D_7 + 12 D_8 + \tilde{D}_{13}  ], \nonumber \\ 
    D_{3P2} &= \frac{-2 \pi}{15} [ 40 D_1 + 40 D_2 +40  D_3 +40 D_4 + 10 D_5 + 10 D_6 + 4 D_7 + 4 D_8 - \tilde{D}_{13}  ], \nonumber \\
    D_{3P0} &= \frac{-4 \pi}{3} [ 4 D_1 + 4 D_2 +4  D_3 +4 D_4 - 2 D_5 - 2 D_6 - 8 D_7 - 8 D_8 - \tilde{D}_{13}  ], \nonumber \\ 
    D_{1D2} &= \frac{8 \pi}{15} [ 4 D_1 + 4 D_2 - 12 D_3 -12 D_4 -4 D_7 -4 D_8 -  \tilde{D}_{11} -  \tilde{D}_{12} + \tilde{D}_{13} + \tilde{D}_{15}  ], \nonumber \\ 
    D_{3D2} &= \frac{4 \pi}{15} [ 8 D_1 - 24 D_2 + 8 D_3 - 24 D_4 - 2 D_5 + 6 D_6 + 12 D_7 - 36 D_8 - 2 \tilde{D}_{11} + 6  \tilde{D}_{12} - 3 \tilde{D}_{13} + 4 \tilde{D}_{15}  ], \nonumber \\ 
    D_{3D1} &= \frac{4 \pi}{45} [ 24 D_1 - 72 D_2 + 24 D_3 - 72 D_4 - 18 D_5 + 54 D_6 - 20 D_7 + 60 D_8 - 6 \tilde{D}_{11} + 18  \tilde{D}_{12} + 5 \tilde{D}_{13} - 16 \tilde{D}_{15}  ], \nonumber \\
    D_{3D3} &= \frac{8 \pi}{15} [ 4 D_1 - 12 D_2 + 4 D_3 - 12 D_4 + 2 D_5 - 6 D_6 -  \tilde{D}_{11} + 3  \tilde{D}_{12} - \tilde{D}_{15}  ], \nonumber \\ 
    D^1_{3D1-3S1} &= \frac{2 \sqrt{2} \pi}{9} [ 28 D_7 - 84 D_8 - \tilde{D}_{13} - 4  \tilde{D}_{15}   ], \nonumber \\ 
    D^2_{3D1-3S1} &= \frac{2 \sqrt{2} \pi}{3} [ 4 D_7 - 12 D_8 + \tilde{D}_{13}], \nonumber \\ 
    D_{3F2-3P2} &= -\frac{4 \sqrt{6} \pi}{15} [ 4 D_7 + 4 D_8 - \tilde{D}_{13}].
\end{align}

\end{widetext}

\section{Testing different sets of nonlocal operators}
\label{sec:Different_nonlocal_sets}

In this work, we have chosen a set of four nonlocal operators specified in Eq.~\eqref{eq:nonlocal_ops}, leading to the construction of the interactions N$^3$LO$_{\rm LA}$-09 to N$^3$LO$_{\rm LA}$-06. 
However, as discussed in Sec.~\ref{subsec:contacts}, other choices for these nonlocal operators are possible due to FRF and the freedom in choosing the parameters of the UT. 
In this appendix, we will briefly investigate four other possible sets of nonlocal operators, with the cutoff fixed at $R_0=0.6$~fm, and compare them with our N$^3$LO$_{\rm LA}$-06 interaction.

The alternative sets, Set 1 to Set 4, are defined by the choice of 4 nonlocal operators as follows:
\begin{align}
V_{\text{Set 1}}&= 
\Tilde{D}_{9}\, \fet q^2 \fet k^2  + \Tilde{D}_{10}\, \fet q^2 \fet k^2 \fet \tau_1\cdot \fet\tau_2 \nonumber \\
&\quad +\Tilde{D}_{13}\, \fet k^2 \, \fet \sigma_1\cdot \fet q \, \fet\sigma_2 \cdot \fet q  \nonumber \\
&\quad +\Tilde{D}_{15}\, (\fet \sigma_1 \cdot \fet L) (\fet \sigma_2 \cdot \fet L)\,, \label{eq:nonloc_1} \\[2mm]
V_{\text{Set 2}}&= 
\Tilde{D}_{9}\, \fet q^2 \fet k^2  + \Tilde{D}_{12}\, \fet L^2 \fet \tau_1\cdot \fet\tau_2 \nonumber \\
&\quad +\Tilde{D}_{13}\, \fet k^2 \, \fet \sigma_1\cdot \fet q \, \fet\sigma_2 \cdot \fet q  \nonumber \\
&\quad +\Tilde{D}_{15}\, (\fet \sigma_1 \cdot \fet L) (\fet \sigma_2 \cdot \fet L)\,, \\[2mm]
V_{\text{Set 3}}&= 
\Tilde{D}_{10}\, \fet q^2 \fet k^2 \fet \tau_1\cdot \fet\tau_2  + \Tilde{D}_{11}\, \fet L^2  \nonumber \\
&\quad +\Tilde{D}_{13}\, \fet k^2 \, \fet \sigma_1\cdot \fet q \, \fet\sigma_2 \cdot \fet q  \nonumber \\
&\quad +\Tilde{D}_{15}\, (\fet \sigma_1 \cdot \fet L) (\fet \sigma_2 \cdot \fet L)\,, \\[2mm]
V_{\text{Set 4}}&= 
\Tilde{D}_{11}\, \fet L^2  + \Tilde{D}_{12}\, \fet L^2 \fet \tau_1\cdot \fet\tau_2 \nonumber \\
&\quad +\Tilde{D}_{14}\, \fet k^2 \, \fet \sigma_1\cdot \fet q \, \fet\sigma_2 \cdot \fet q  \fet \tau_1\cdot \fet\tau_2 \nonumber \\
&\quad +\Tilde{D}_{15}\, (\fet \sigma_1 \cdot \fet L) (\fet \sigma_2 \cdot \fet L).
\label{eq:nonloc_4}
\end{align}

\begin{figure*}
    \includegraphics[scale=0.5]{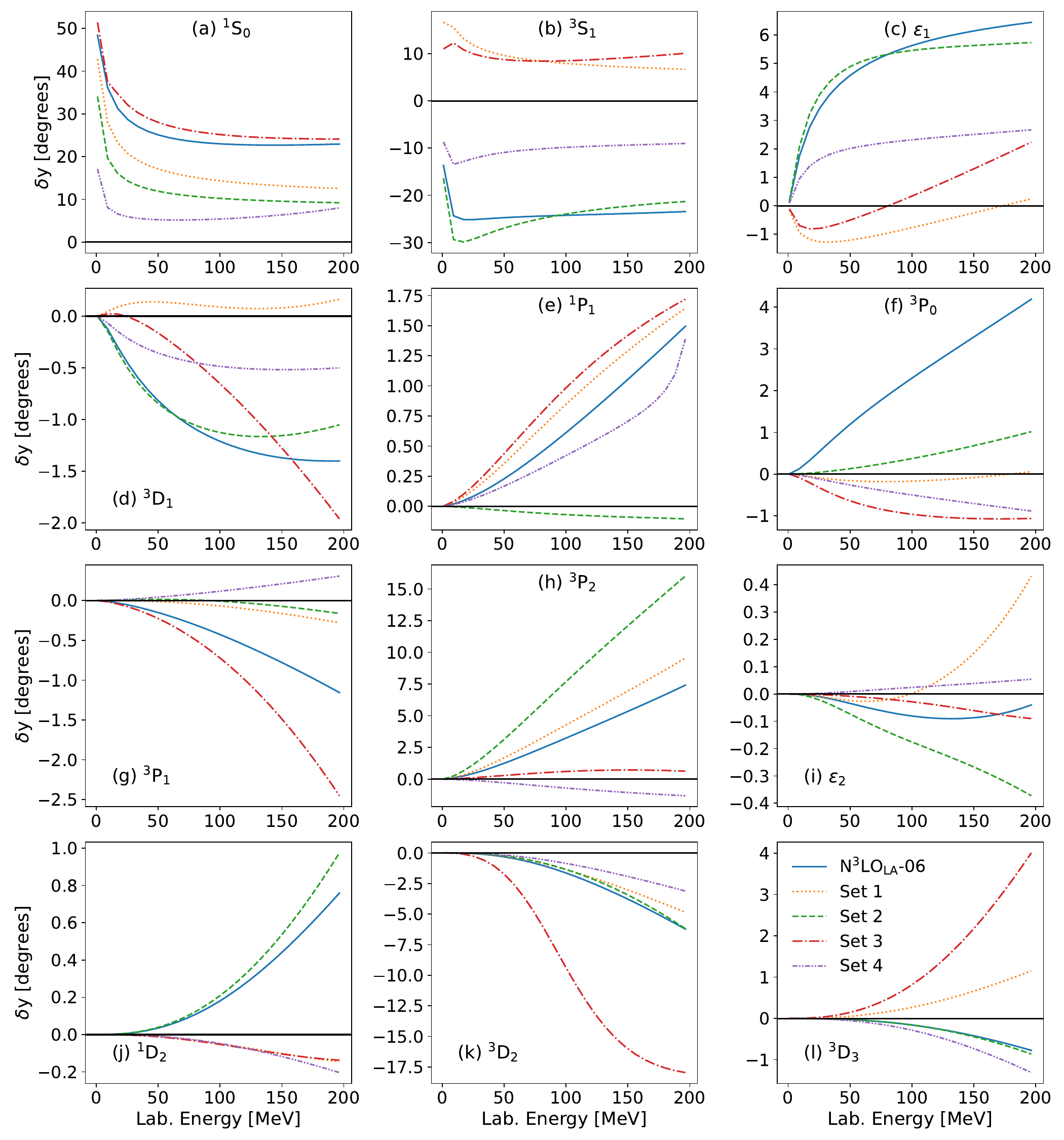}
    \caption{Differences between the phase shifts obtained using the full interaction and the predictions with all nonlocal parts set to zero. We show results for the N$^3$LO$_{\rm LA}$-06 interaction as well as the interactions defined in Eqs.~\eqref{eq:nonloc_1} to~\eqref{eq:nonloc_4}, i.e., Set 1 to Set 4.}
    \label{fig:different_non_local_sets}
\end{figure*}

An important consideration in choosing the operator set is regarding the perturbativeness of the nonlocal pieces, which is important for their application in QMC calculations.
In Fig.~\ref{fig:different_non_local_sets}, we show the differences between the phase shifts in several partial waves for these interactions and the predictions with all nonlocal parts set to zero.
By examining these differences, we can judge the relative strength of the nonlocal terms, i.e., the extent to which the full solution is determined by the local operators alone. 
While the question of the perturbativeness of the nonlocal operators depends on the observable in question, this phase shift analysis could indicate the sets of nonlocal operators that would be best suited for a perturbative treatment in many-body calculations. 
Generally, the performance of all sets are comparable, indicating that the choice of nonlocal operators made in this work, Eq.~\eqref{eq:nonlocal_ops}, might present no particular roadblock for QMC applications of nuclei and nuclear matter.

\section{Impact of the SFR cutoff}
\label{sec:SFR}

\begin{figure*}
    \includegraphics[scale=0.5]{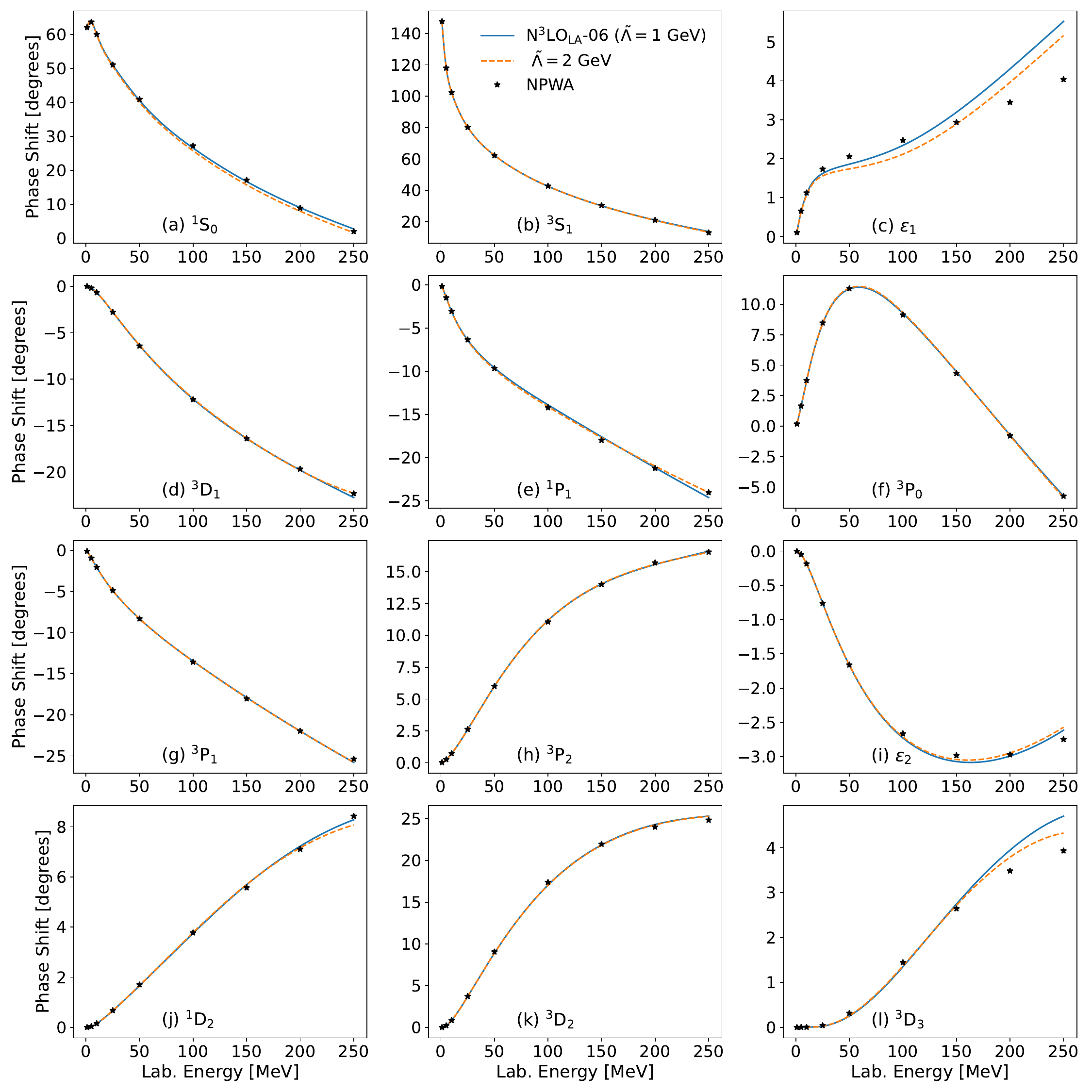}
    \caption{Impact of changing the SFR cutoff $\tilde{\Lambda}$. The solid blue line corresponds to our N$^3$LO$_{\rm LA}$-06 interaction with $\tilde{\Lambda}=1$\,GeV and the dashed orange curve is obtained with the same interaction but with $\tilde{\Lambda}=2$\,GeV.}
    \label{fig:0.6_SFR_2GeV}
\end{figure*}

Next, we study the impact of changing the SFR cutoff, see Sec.~\ref{sec:pions}. 
In Fig.~\ref{fig:0.6_SFR_2GeV}, we show the phase shifts in several partial waves for our N$^3$LO$_{\rm LA}$-06 interaction with $\tilde{\Lambda}=1$ GeV and compare this with results for the interaction with $\tilde{\Lambda}=2$ GeV using the same fit protocol. 
Since the two cases are almost identical, we conclude that the SFR cutoff does not significantly impact our results. This is consistent with what was observed for the local interactions at N$^2$LO~\cite{Gezerlis:2013ipa,Gezerlis:2014zia}.

\section{Comparison of the BUQEYE and EKM prescriptions for EFT truncation errors}
\label{sec:Buqeye_vs_EKM}

We compare the EKM Eq.~\eqref{eq:EKM} and BUQEYE Eq.~\eqref{eq:buqeye} models for the EFT truncation errors at \NNNLO for the cutoff $R_0 = 0.6$~fm. 
As shown in Eq.~\eqref{eq:buqeye}, the authors of Ref.~\cite{Wesolowski:2018lzj} obtain a Gaussian distribution for the EFT truncation errors. 
In this appendix, we will use the updated work of Ref.~\cite{Melendez:2019izc}. 
In this approach, the BUQEYE prescription for the posterior on the observable calculated at order $k$, $X^k$ is given by the Student-t distribution, 
\begin{equation}
    P(X) \propto t_\nu \bigg[X^k, X_\text{ref}^2 \frac{Q^{2(k+1)}}{1-Q^2} \tau^2 \bigg]\,,
    \label{eq:b_pdf}
\end{equation}
where $\nu$ and $\tau$ are hyperparameters given as,
\begin{align}
    \nu &= \nu_0 + n_c \nonumber \\
    \nu \tau^2 &= \nu_0 \tau_0^2 + \fet{c_k}^2\,, \nonumber \\
\end{align}
where we make the uninformed choices $\nu_0 = \tau_0 = 1$, $\fet{c_k}$ is the vector containing the expansion coefficients $c_n$, and $n_c$ is the number of expansion coefficients. 

\begin{figure*}
    \centering
    \includegraphics[scale=0.62]{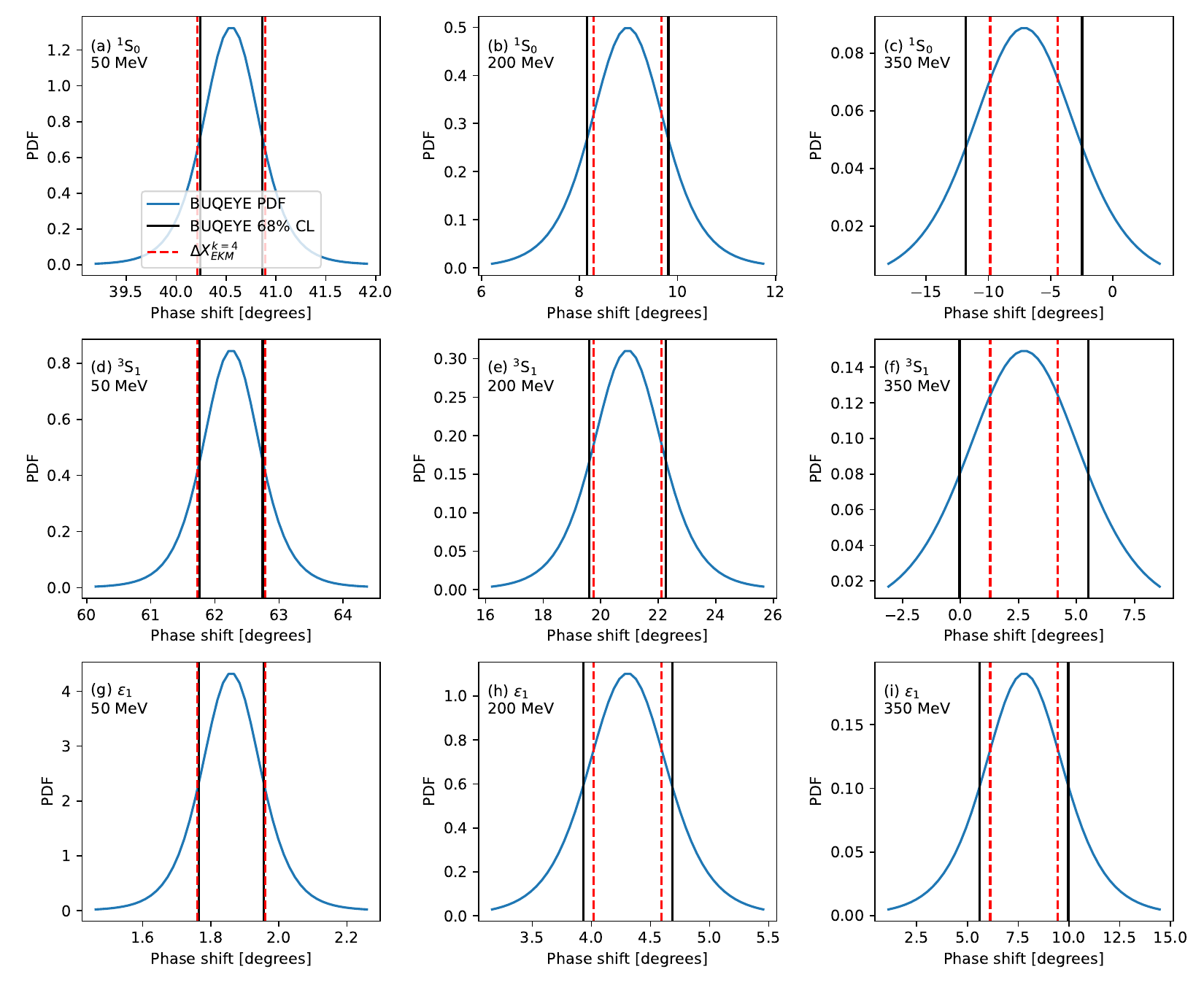}
    \caption{Comparison of the BUQEYE estimate of the EFT truncation uncertainties with the EKM estimate in different $S$-wave channels. 
    This serves as input to our Bayesian analysis—see Eq.~\eqref{eq:L}—and similar error estimates will result in similar Bayesian fit posteriors.
    The blue distribution is obtained using Eq.~\eqref{eq:b_pdf} and the solid black lines are the corresponding 68\% CL. 
    The EKM uncertainty, i.e., Eq.~\eqref{eq:EKM} with $k=4$, is shown as dashed red lines. 
    Different panels correspond to different partial waves and laboratory energies as written in the upper left of each panel.}
    \label{fig:buqeye_vs_EKM_S}
\end{figure*}

\begin{figure*}
    \centering
    \includegraphics[scale=0.62]{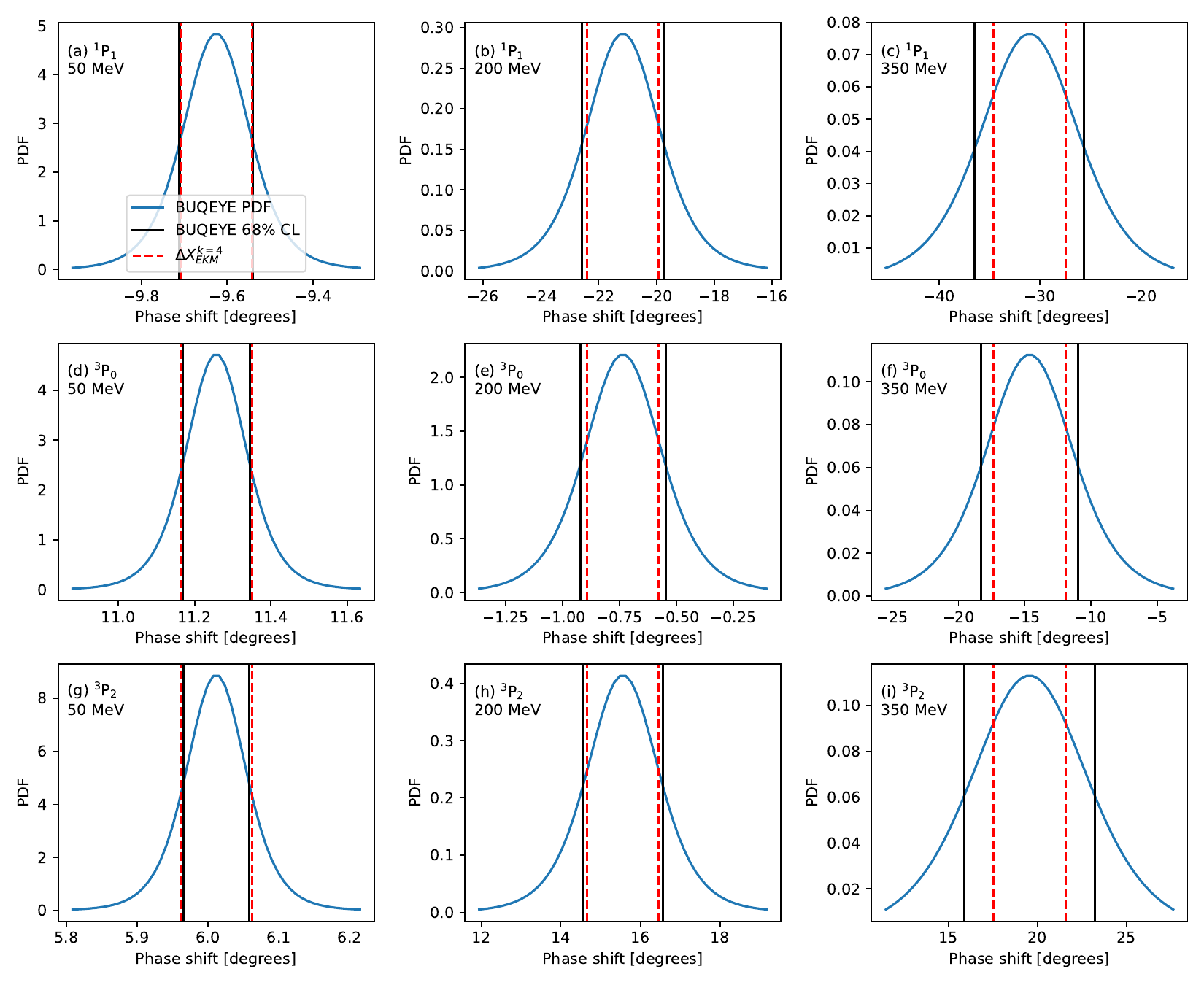}
    \caption{Same as Fig.~\ref{fig:buqeye_vs_EKM_S} but for $P$ waves.}
    \label{fig:buqeye_vs_EKM_P}
\end{figure*}

\begin{figure*}
    \centering
    \includegraphics[scale=0.62]{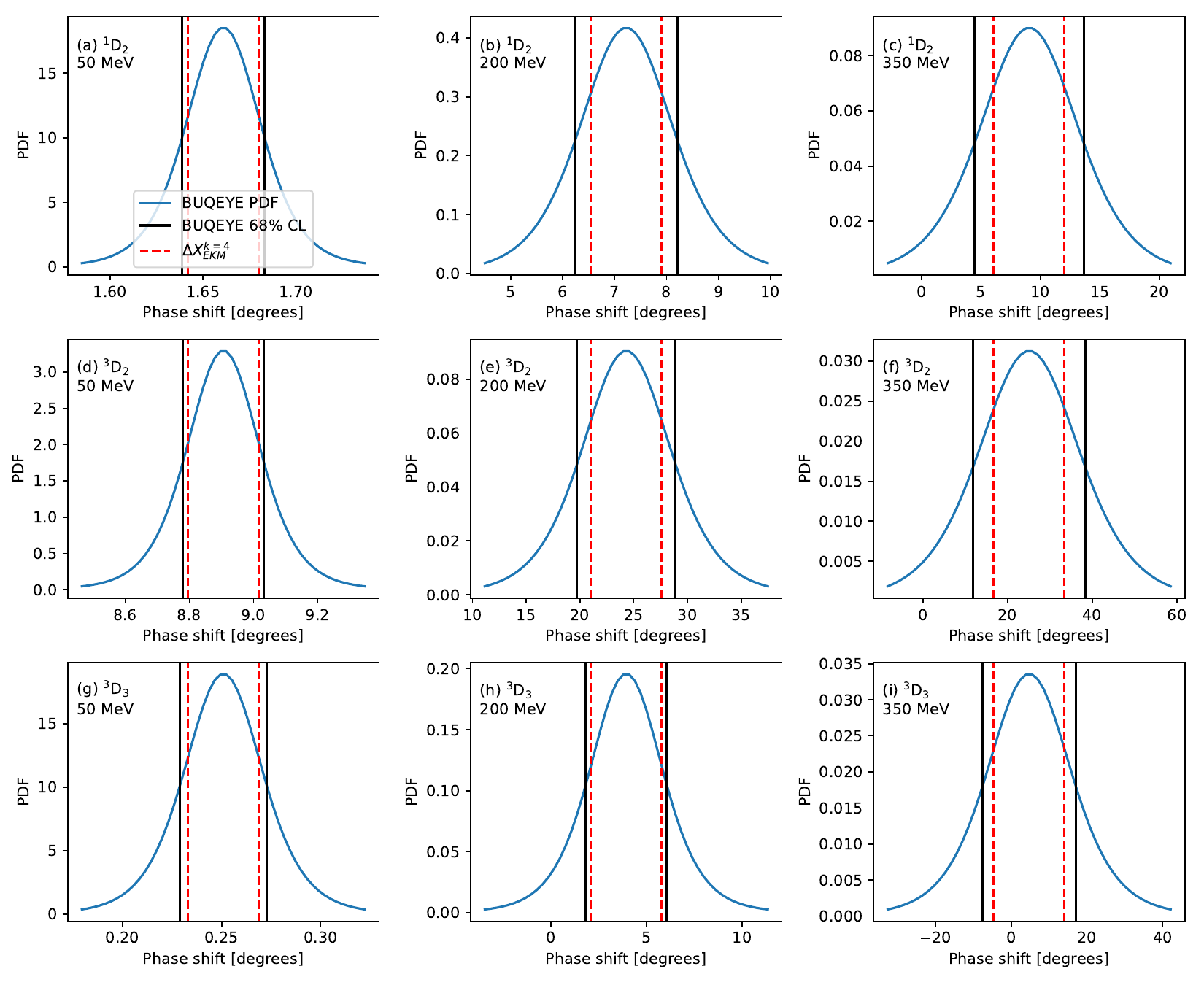}
    \caption{Same as Fig.~\ref{fig:buqeye_vs_EKM_S} but for $D$ waves.}
    \label{fig:buqeye_vs_EKM_D}
\end{figure*}

In Figs.~\ref{fig:buqeye_vs_EKM_S},~\ref{fig:buqeye_vs_EKM_P}, and~\ref{fig:buqeye_vs_EKM_D}, the posterior distribution function as given in Eq.~\eqref{eq:b_pdf} is shown in blue and the corresponding 68\% CL is shown as black vertical lines. 
In each figure, the rows correspond to different partial waves and the columns correspond to different laboratory energies. 
We also show the EKM estimate obtained using Eq.~\eqref{eq:EKM}, with $k=4$, as red dashed lines. 
We find that, in all partial waves, the two methods are almost identical at lower energies, with the EKM estimate producing smaller bands at higher energies. 
This is due to the fact that the EKM estimate takes into account only the first omitted term in the EFT expansion, whereas the BUQEYE prescription sums all the higher-order terms as well. 
Note, however, that at $E = 200$~MeV the two estimates still provide similar results.
Given that we use the EKM estimate as the variance of our likelihood function only until $E_\text{max} = 250$~MeV, we conclude that using Eq.~\eqref{eq:L}  instead of Eq.~\eqref{eq:buqeye_L} as our likelihood function would not significantly impact the results of our Bayesian fits. 

\section{Operator LECs at $\mathrm{LO}$, \NLO, and \NNLO}
\label{sec:LECs_lower_orders}

Finally, we give the operator LECs that we have obtained from our phase shift analyses at LO, \NLO, and \NNLO. 
The LECs are quoted in Tables~\ref{tab:LEC_N2LO},~\ref{tab:LEC_NLO}, and~\ref{tab:LEC_LO}.

\begin{table*}[t]
\centering
\tabcolsep=0.5cm
\def\arraystretch{1.5}
\begin{tabular}{c|cccc|cccc}
\hline
LEC & \multicolumn{4}{c|}{Maximum posterior estimate} & \multicolumn{4}{c}{Least-squares fit}  \\
\hline
                & $0.9$~fm & $0.8$~fm & $0.7$~fm & $0.6$~fm & $0.9$~fm & $0.8$~fm & $0.7$~fm & $0.6$~fm\\
\hline
$C_S$~[fm$^2$]       & $0.666$ & $1.971$ & $4.196$ & $8.641$ & $1.299$ & $2.679$ & $4.965$ & $9.157$ \\
$C_T$~[fm$^2$]      & $0.467$ & $0.508$ & $0.699$ & $1.426$ & $0.684$ & $0.749$ & $0.951$ & $1.607$ \\
$C_1$~[fm$^4$]     & $-0.001$ & $-0.05$ & $-0.097$ & $-0.211$ & $-0.057$ & $-0.04$ & $-0.072$ & $-0.172$ \\
$C_2$~[fm$^4$]      & $-0.034$ & $0.024$ & $0.035$ & $0.072$ & $0.127$ & $0.096$ & $0.072$ & $0.064$ \\
$C_3$~[fm$^4$]      & $-0.093$ & $-0.095$ & $-0.087$ & $-0.096$ & $-0.092$ & $-0.086$ & $-0.081$ & $-0.08$ \\
$C_4$~[fm$^4$]      & $0.06$ & $0.073$ & $0.103$ & $0.156$ & $0.112$ & $0.108$ & $0.122$ & $0.151$ \\
$C_5$~[fm$^4$]      & $-1.948$ & $-1.991$ & $-2.23$ & $-2.897$ & $-2.158$ & $-2.168$ & $-2.354$ & $-2.948$ \\
$C_6$~[fm$^4$]      & $0.303$ & $0.208$ & $0.18$ & $0.196$ & $0.287$ & $0.23$ & $0.2$ & $0.202$ \\
$C_7$~[fm$^4$]      & $-0.479$ & $-0.353$ & $-0.308$ & $-0.356$ & $-0.507$ & $-0.406$ & $-0.353$ & $-0.361$ \\
\hline
\end{tabular}
\caption{Same as Table~\ref{tab:LEC_N3LO} but at \NNLO.}
\label{tab:LEC_N2LO}
\end{table*}

\begin{table*}[t]
\centering
\tabcolsep=0.5cm
\def\arraystretch{1.5}
\begin{tabular}{c|cccc|cccc}
\hline
LEC & \multicolumn{4}{c|}{Maximum posterior estimate} & \multicolumn{4}{c}{Least-squares fit}  \\
\hline
                & $0.9$~fm & $0.8$~fm & $0.7$~fm & $0.6$~fm & $0.9$~fm & $0.8$~fm & $0.7$~fm & $0.6$~fm\\
\hline
$C_S$~[fm$^2$]       & $0.083$ & $0.95$ & $2.576$ & $7.954$ & $0.043$ & $1.503$ & $4.952$ & $17.72$ \\
$C_T$~[fm$^2$]      & $0.326$ & $0.662$ & $1.045$ & $0.329$ & $0.881$ & $1.082$ & $1.946$ & $5.068$
 \\
$C_1$~[fm$^4$]     & $0.125$ & $0.598$ & $0.59$ & $0.887$ & $0.256$ & $0.483$ & $0.553$ & $0.848$
 \\
$C_2$~[fm$^4$]      & $0.095$ & $0.173$ & $0.091$ & $-0.017$ & $0.162$ & $0.24$ & $0.277$ & $0.42$\\
$C_3$~[fm$^4$]      & $0.109$ & $-0.242$ & $-0.11$ & $-0.116$ & $-0.07$ & $-0.169$ & $-0.097$ & $-0.137$
 \\
$C_4$~[fm$^4$]      & $0.148$ & $0.037$ & $0.043$ & $-0.028$ & $0.128$ & $0.066$ & $0.123$ & $0.275$
\\
$C_5$~[fm$^4$]      & $-1.876$ & $-1.948$ & $-2.103$ & $-2.386$ & $-2.22$ & $-2.304$ & $-2.659$ & $-3.713$
 \\
$C_6$~[fm$^4$]      & $0.367$ & $0.321$ & $0.261$ & $0.132$ & $0.355$ & $0.337$ & $0.393$ & $0.63$
\\
$C_7$~[fm$^4$]    & $-0.472$ & $-0.369$ & $-0.277$ & $-0.125$ & $-0.509$ & $-0.432$ & $-0.442$ & $-0.643$
 \\
\hline
\end{tabular}
\caption{Same as Table~\ref{tab:LEC_N3LO} but at \NLO.}
\label{tab:LEC_NLO}
\end{table*}

\begin{table*}[t!]
\centering
\tabcolsep=0.5cm
\def\arraystretch{1.5}
\begin{tabular}{c|cccc|cccc}
\hline
LEC & \multicolumn{4}{c|}{Maximum posterior estimate} & \multicolumn{4}{c}{Least-squares fit}  \\
\hline
                & $0.9$~fm & $0.8$~fm & $0.7$~fm & $0.6$~fm & $0.9$~fm & $0.8$~fm & $0.7$~fm & $0.6$~fm\\
\hline
$C_S$~[fm$^2$]       & $-2.527$ & $-1.983$ & $-1.38$ & $-0.634$ & $-2.627$ & $-2.074$ & $-1.464$ & $-0.712$ \\
$C_T$~[fm$^2$]      & $-0.061$ & $0.029$ & $0.141$ & $0.302$ & $-0.08$ & $0.013$ & $0.126$ & $0.287$
\\
\hline
\end{tabular}
\caption{Same as Table~\ref{tab:LEC_N3LO} but at LO.}
\label{tab:LEC_LO}
\end{table*}

\bibliography{biblio}

\end{document}